\documentclass[acmsmall,screen]{acmart}

\AtBeginDocument{%
  }

\setcopyright{acmcopyright}
\copyrightyear{2018}
\acmYear{2018}
\acmDOI{XXXXXXX.XXXXXXX}

\usepackage{amssymb}
\usepackage{amsmath, amsfonts}

\usepackage{graphicx}
\usepackage{textcomp}
\usepackage{xcolor}
\usepackage{hyperref}
\usepackage{cleveref}
\usepackage{xspace}
\usepackage{paralist}
\usepackage{multirow}
\usepackage{subcaption}
\usepackage{tikz}
\usepackage{courier}
\usepackage{listings} 
\usepackage{array}
\usepackage{graphics}
\usepackage{lipsum}
\usepackage{arydshln}
\usepackage{todonotes}
\newcommand{\etal}{\emph{et al.}\xspace}

\newcommand{\ie}{\emph{i.e.},\xspace}
\newcommand{\eg}{\emph{e.g.},\xspace}
\newcommand{\company}{{Microsoft}\xspace}

\definecolor{Gray}{gray}{0.3}
\tikzstyle{mybox} = [draw=black, very thick, rectangle, rounded corners, inner ysep=5pt, inner xsep=5pt, fill=gray!20]
\newcommand{\xyz}[2]{
    \smallskip
    \noindent
    \begin{tikzpicture}
        \node [mybox] (box){%
        \centering
        \begin{minipage}{.97\textwidth}
        \fontsize{8.8}{10}\selectfont
        \textbf{RQ #1}. #2
        \end{minipage}
        };
    \end{tikzpicture}%
}

\begin{document}

\title{Studying LLM Performance on Closed- and Open-source Data}


\author{Toufique Ahmed}
\affiliation{%
 \institution{University of California, Davis}
 \city{Davis}
 \state{California}
 \country{USA}}

\author{Christian Bird}
\affiliation{%
 \institution{Microsoft Research}
 \city{Redmond}
 \state{Washington}
 \country{USA}}

\author{Premkumar Devanbu}
\affiliation{%
 \institution{University of California, Davis}
 \city{Davis}
 \state{California}
 \country{USA}}

\author{Saikat Chakraborty}
\affiliation{%
 \institution{Microsoft Research}
 \city{Redmond}
 \state{Washington}
 \country{USA}}

\renewcommand{\shortauthors}{Ahmed et al.}

\begin{abstract}
Large Language models (LLMs) are finding wide use in software engineering practice. These models are extremely data-hungry, and are largely trained on open-source (OSS) code distributed with permissive licenses. In terms of actual use however, a great deal of software development still occurs in the for-profit/proprietary sphere, where the code under development is not, and never has been, in the public domain; thus,  many developers, do their work, and use LLMs, in settings where the models may not be as familiar with the code under development. In such settings, do LLMs work as well as they do for OSS code? If not, what are the differences? When performance differs, what are the possible causes, and are there work-arounds? In this paper, we examine this issue using proprietary, closed-source software data from \company, where most proprietary code is in C\# and C++. We find that performance for C\# changes little from OSS $\rightarrow$ proprietary code, but does significantly reduce for C++; we find that this difference is attributable to differences in identifiers. We also find that some  performance degradation, in some cases, can be ameliorated efficiently by in-context learning. 
\end{abstract}

\keywords{GPT 3.x, LLMs, Generalization}



\maketitle

\section{Introduction}
Large language models (LLMs), like the ones used by CoPilot~\cite{ziegler2022productivity} are generally trained on very large source code corpora. LLMs are used in a wide variety of tasks, most notably for code-completion. Published
research suggests that code-completion tools based on these models~\cite{murali2023codecompose,ziegler2022productivity,tabachnyk2022ml} offer
substantial improvements in software productivity.

Some of these models~\cite{murali2023codecompose,tabachnyk2022ml} have been built primarily for non public-use: they are trained on proprietary code, specific to the industrial organization within which they are designed, built, and deployed. However, very popular models, such as the OpenAI Codex series~\cite{ziegler2022productivity}, are trained  \emph{almost exclusively} on open-source data. 
Despite being trained exclusively on open-source data, Codex- style models are still widely used in both closed-source (commercial) and open-source settings. Some organizations, such as Facebook~\cite{murali2023codecompose} and Google~\cite{tabachnyk2022ml} have reported on the performance of \emph{closed} models, 
trained on \emph{commercial, company-specific data}. However, the performance of models like Codex, which are trained on \emph{publicly available, open-source data}, when used on \emph{commercial}, 
non-open source code is not yet fully understood. This is the focus of our paper.

Why would this be a concern? Wouldn't these models just work? Aren't software projects sufficiently similar to each other, because of Software Naturalness~\cite{hindle2012naturalness}? While the question isn't directly addressed in the literature, prior work points at a range of different conclusions, on different tasks. Consider the task of cross-project defect prediction: \textit{can defect predictors trained on one project work on another?} This could be viewed as a fairly simple task, compared to the range of tasks undertaken by language models today. Early work suggests that straightforward cross-project prediction \textit{does not work}~\cite{zimmermann2009cross} while
more sophisticated, targeted methods do~\cite{he2013learning}. However, these latter approaches~\cite{hosseini2017systematic,pal2022cross} generally require detailed knowledge of the differences between training and test data, and consequent adaptation of ML methods in use; in general, such approaches are not useful in the settings where general-purpose large language models like Codex are used. 

More specifically for language models, early work by Tu \emph{et al.}~\cite{tu2014localness} showed that language models \emph{needed} to be aware of explicit locality at the very fine-grained level of \emph{files!} This issue arises from the tendency of programmers to invent file-specific identifiers and usage patterns (\eg method calls, idioms, etc); if a language model ignored these details, the performance suffered. This work was further elaborated by Hellendoorn \etal~\cite{hellendoorn2017deep}, who showed that locality at multiple levels (files, directories, packages, projects) existed, and could be exploited. These above works used discrete (n-gram) LLMs. With the advent of sub-tokenization~\cite{karampatsis2020big}, where long, newly-invented words could be split up into sequences of ``sub-tokens'',  deep-learning models have been able to deal better with localization phenomena such as identifier names. 

Very recent work using the few-shotting ability of  large language models, \textit{viz.,} also provides evidence suggesting that LLMs exhibit project-specific changes in behaviour. For example Ahmed \etal~\cite{ahmed2022few} found that providing project-specific exemplars in a few-shot context provided better performance than others chosen at random. The same authors also report evidence suggesting that identifier usage plays a strong role in language model performance~\cite{ahmed2022multilingual}; identifier names
are very project-specific, and could be expected to differ from setting to setting. Other work~\cite{nashid2023retrieval} suggests that finding  few-shot exemplars most relevant to
the given input also helps generate better outputs. This raises the question, when LLMs work with non-open source data, is their performance the same? Or different from working with closed-source data?

Commercial projects with non-public source code,  do have special features. Projects in large software organizations could make use of specific naming standards, coding conventions, proprietary APIs (and associated usage patterns), which could all potentially make their code substantially different from the publicly available code that LLMs were trained on. Thus, one might expect that models trained exclusively on publicly available code would perform differently on non-public code, both on the basic code-completion tasks, as well as other tasks (\ie code summarization, code generation). To explore this further, we conducted a study of the publicly available OpenAI CoPilot models in this context. The research was conducted as a collaboration between University-based researchers, and employees of a large software firm (\company) with a significant and diverse body of software assets. 

\vspace{.2cm}

We make the following contributions: 
\begin{itemize}
    \item We compare the \emph{code completion} performance of the CoPilot models on publicly-available code, and non-public, proprietary code. We find very little or no difference for C\#, but statistically significant difference for C++. 
    \item We compare the \emph{Few-shot} performance of code summarization and code generation between OSS and closed-source samples, and like code completion, find less differences for C\# but higher differences for C++.  
    \item We investigate many potential \emph{causal reasons} behind these differences.  While results are not conclusive, they do offer directions for additional studies. 
\item We also show and quantify how using few-shot samples from both OSS and closed-source samples can enhance the models' performance on closed-source data.
\end{itemize}
\section{Background \& Motivation}

LLMs-based coding assistants like Copilot are widely used. Even though LLMs sometimes generate buggy programs~\cite{jesse2023large}, developers like to use LLMs for different tasks and view them positively~\cite{bird2022taking}. Ziegler et al.~\cite{ziegler2022productivity} asked users of GitHub Copilot about its impact on their productivity and found evidence of their (very positive) views, directly in measurable user data. Several surveys have explained how Large Language Models (LLMs) have significantly impacted numerous domains, including Software Engineering (SE), and many recent publications have explored LLMs applied to various SE tasks~\cite{fan2023large,hou2023large}. In this work, we seek a better understanding how LLMs work, given the increasing diversity of the use base, especially with respect to open- and closed-source systems.

\subsection{Open-source vs. Closed-source}

Closed-source and open-source code differ fundamentally in their development style, accessibility, and usage characteristics.  Certainly, this distinction is rather coarse, since there are many styles of development within both OSS and closed-source.  Still, it is useful to outline some general characteristics that differentiate OSS and closed-source.

Differences between Open-source software (OSS) and closed-source software~\cite{paulson2004empirical} arise from differing development models, objectives, and operational frameworks. 
Open source development usually involves a collaborative, community-driven approach, leading to different coding styles whereas closed-source projects, developed by specific teams within a company, tend to have more uniform coding practices. 
The transparency inherent in OSS allows for extensive public scrutiny and improvement, potentially enhancing code quality and security; this differs from closed-source software, which relies almost entirely on internal review processes, but  may also have other resources (e.g., enhanced testing infrastructures) for quality assurance. 
OSS systems are also more likely to mutually depend on other open-source systems, thus fostering an ecosystem of reusable code, while closed-source projects may lean towards proprietary solutions. 
As open-source projects rely on newcomers, they often feature comprehensive documentation to aid community contributions and they also evolve rapidly based on user and community feedback. 
In contrast, closed-source projects prioritize business requirements and often follow a top-down approach to feature development. 
These fundamental differences in collaboration, transparency, innovation, dependencies, documentation, and user feedback mechanisms contribute to differences in the form and functionality of open-source versus closed-source code.

Evaluating potentially different performance of LLMs across code from the two paradigms is important for several reasons. 
First, by shedding light on the performance differences between use on OSS code and unseen industrial code, our results can help commercial developers and stakeholders make decisions concerning the value of 
LLM-powered tools in their development practices.
Second, by understanding how LLMs perform on different types of code, researchers can gain insights
on model generalization and robustness. Looking forward, by understanding the \emph{causes} of any performance discrepancies observed between open- and closed-source code, researchers could develop models with better performance across diverse data sources and a wider range of real-world applications. 

\subsection{Why Study Multiple Languages?}
An integral yet often overlooked dimension in the realm of large language models' performance in software tasks is the programming language in use. Earlier studies~\cite{casalnuovo2019studying} suggest LLM performance may vary across languages, even in open-source contexts. In addition, 
different languages, with their unique syntax, standard libraries, and ecosystems, could present \emph{differing} challenges and patterns when transitioning from open to closed source contexts, challenges that are specific to each language. Several reasons compound the importance of investigating multiple languages:

\vspace{.2cm}

\noindent{\underline{\emph{Language Ecosystems:}}} Open-source communities for certain languages may be more active and diverse than others~\cite{githubProgrammingLanguages}, 
leading to richer and more varied data for training. 
This can influence how well an LLM generalizes to closed-source projects in that particular language.

\vspace{.2cm}
  
\noindent{\underline{\emph{Usage Scenarios:}}} Certain languages may be predominantly used in specific sectors~\cite{thinkfulCodingLanguages}. For example, while C\# might be used in a wide range application programming in both open and closed environments~\cite{wikipediaListSharp}, other languages like C++ might be more popular in Systems and Embedded Software settings~\cite{thinkfulCodingLanguages}.

\vspace{.2cm}
    
\noindent{\underline{\emph{Language Evolution:}}} The pace and nature of language evolution can differ. Some languages might see rapid changes in closed-source industry settings due to proprietary advancements, whereas others might evolve more sedately, and transaprently, in open-source settings.

\vspace{.2cm}
    
\noindent{\underline{\emph{Coding Conventions:}}} Different languages might have distinct coding conventions and styles in open-source and closed-source environments. These conventions can affect the predictions and utility of code completion tools.

\vspace{.2cm}
      
\noindent{\underline{\emph{Library and Framework Usage:}}} Open and closed-source projects might use different sets of libraries and frameworks, which can be more pronounced in certain languages. An LLM trained predominantly on open-source data might be unfamiliar with proprietary libraries commonly used in closed-source projects of a particular language.




\vspace{.2cm}

Given these potential variances, it becomes imperative to scrutinize the performance of LLMs across multiple languages. We selected two languages that differ in most of the above categories. Note that our choices are also restricted by the languages widely used at \company.


\section{Research Questions}

In the first research question, we will investigate how LLM performs on open- and closed-source data. As mentioned earlier, prior studies have reached different conclusions, but those findings may not apply to LLM today, given the changes in the model architecture and the significant increase in the amount of training data. Our experiments are done without any fine-tuning or weight-updates on the model. 

For this research, we will collect an equal number of testing samples from both \company and open-source datasets and compare the performance of LLM. It is essential to note that data from \company is proprietary, and 
has never been exposed in any setting from whence the training data for OpenAI's models could be collected. Thus we can compare the performance of these LLMs on publicly available code from sources where the training was collected, and sources inaccessible
to OpenAI's training data collection methods.

\xyz{1}{Do LLMs perform differently with open-source and closed-source data?}

We  also study whether open-source samples can improve the performance of closed-source data, 
an \emph{in-context} learning~\cite{dong2022survey} setting. We present exemplars (as detailed in Section~\ref{rq2}), in a few-shot context, for specific tasks. Recent studies~\cite{nashid2023retrieval,ahmed2023improving} have found that choosing exemplars using a retrieval algorithm can help performance.
In a retrieval-based approach, we need a pool of samples, for which closed-source samples from within the organization may be limited; but many examples are available from open-source projects. Therefore, a natural question arises: 


\xyz{2}{Can the performance on closed-source projects be improved using few-shot samples from open-source projects?}

Our findings relating to RQ1 (Section~\ref{rq1}), suggest that Language Models (LLMs) exhibit consistent or similar performance on both open-source and closed-source data for C\#; however, this trend does not hold for C++: the models perform poorly across all tasks related to closed-source C++. Our research investigates the potential reasons behind these differences. Our hope is that we could develop a better understanding
for the reasons for model's poor performance, and perhaps address them. 

\xyz{3}{What causes the variance in performance between the models for two different languages?}


\section{Tasks, Dataset, and Models}
In this section, we briefly discuss the Tasks we used to evaluate the LLMs in open and closed-source context as well as and the Data we used.

\subsection{Tasks}
\label{tasks}
To evaluate the model's performance on both open-source and closed-source data, we selected four distinct tasks: token completion, line completion, code summarization, and code generation. All four tasks have been included in the popular CodeXGLUE~\cite{lu2021codexglue} benchmark for software engineering tasks.

\subsubsection{Token Completion}
Here, the goal is to predict the next token in the body of  a function, conditioned on the preceding tokens. We employ Language Models (LLMs) trained on an extensive corpus comprising billions of tokens. The model learns to auto-regressively predict the next token in sequences during training. To assess the model's accuracy, we compared open-source and closed-source projects. We collected uniformly at random 10,000 C\# and 10,000 C++ samples from each category from the corpus.
We randomly selected one token and asked the model to generate that token, given all the preceeding tokens in the sequence.
It is important to note that we employed a zero-shot setting, refraining from leveraging additional information to enhance the model's performance.
We employed an exact match criterion to gauge the model's performance for evaluation purposes. This criterion necessitates an exact match between the predicted and actual tokens. Using this approach, we assessed the model's predictive capabilities without any leniency for partial matches.

\subsubsection{Line Completion}

Line completion is like token completion, but rather than generating a single token, we ask it to generate an entire line. To evaluate the model's performance, we collected uniformly randomly chosen 10,000
function-level samples of both C\# and C++ code, comprising both open-source and closed-source functions. For each sample, we randomly selected a line and tasked the model with generating that line, given all the preceding lines.
In line with the CodeXGLUE benchmark, we utilize two metrics to measure the model's performance: Exact Match (EM) and Edit Similarity (ES). This task better aligns with developers' activities, as they often deal with complete lines rather than individual words when using AI models for automatic completions.

\begin{table}[t]

\centering

\small
{%
\renewcommand{\arraystretch}{1.2}
\begin{tabular}{lllrr}
\hline
\multicolumn{1}{c}{Langiage} & \multicolumn{1}{c}{Task}            & \multicolumn{1}{c}{Category} & \begin{tabular}[c]{@{}c@{}}\# of training/ \\ few-shot samples\end{tabular} & \begin{tabular}[c]{@{}c@{}}\# of test \\ samples\end{tabular} \\ \hline
\multirow{8}{*}{C\#}         & \multirow{2}{*}{Token completion}   & OSS                        & NA                                                                          & 10,000                                                        \\
                             &                                     & Proprietary                & NA                                                                          & 10,000                                                        \\  \cline{2-5}
                             & \multirow{2}{*}{Line completion}    & OSS                        & NA                                                                          & 10,000                                                        \\ 
                             &                                     & Proprietary                & NA                                                                          & 10,000                                                        \\ \cline{2-5}
                             & \multirow{2}{*}{Code summarization} & OSS                        & 31,641                                                                      & 1,000                                                         \\  
                             &                                     & Proprietary                & 5,904                                                                         & 1,000                                                         \\ \cline{2-5}
                             & \multirow{2}{*}{Code Completion}    & OSS                        & 31,641                                                                      & 1,000                                                         \\
                             &                                     & Proprietary                & 5,904                                                                         & 1,000                                                         \\ \hline
\multirow{8}{*}{C++}         & \multirow{2}{*}{Token completion}   & OSS                        & NA                                                                          & 10,000                                                        \\
                             &                                     & Proprietary                & NA                                                                          & 10,000                                                        \\ \cline{2-5}
                             & \multirow{2}{*}{Line completion}    & OSS                        & NA                                                                          & 10,000                                                        \\
                             &                                     & Proprietary                & NA                                                                          & 10,000                                                        \\ \cline{2-5}
                             & \multirow{2}{*}{Code summarization} & OSS                        & 142,195                                                                     & 1,000                                                         \\
                             &                                     & Proprietary                & 6,643                                                                         & 1,000                                                         \\ \cline{2-5}
                             & \multirow{2}{*}{Code Completion}    & OSS                        & 142,195                                                                     & 1,000                                                         \\
                             &                                     & Proprietary                & 6,643                                                                         & 1,000     \\ \hline                                                    
\end{tabular}
}
\vspace{0.05in}
\caption{Dataset for different tasks.}
\vspace{-0.2in}
\label{tbl:dataset}
\end{table}

\subsubsection{Code Summarization}
Developers often use comments to comprehend the specifications and design of the code they are working on, even when comments are only approximate. The task here is to generate a natural language summary of a given piece of code in a specific programming language (PL).
Code summarization has evolved, progressing from template-based approaches~\cite{eddy2013evaluating,haiduc2010supporting,haiduc2010use,sridhara2010towards,rodeghero2014improving} to applying Neural Machine Translation (NMT)~\cite{ahmad2020transformer, leclair2019neural,iyer2016summarizing,hu2018summarizing} and now leveraging large language models (LLMs)~\cite{chen2021evaluating,ahmed2022few}. Many studies have explored the effectiveness of LLMs in the code summarization task. However, the primary objective of this study is not to achieve the best performance on the code summarization task but to demonstrate how LLMs generalize and how their performance varies between open-source and closed-source programs. To accomplish this, we study two recent LLM-based approaches: few-shot learning~\cite{brown2020language,ahmed2022few} and BM25-retrieved few-shot learning~\cite{nashid2023retrieval} which have been applied to Software Engineering tasks. These approaches will be discussed in Section~\ref{rq1-method}.
It is important to note that numerous studies have addressed data quality and evaluation metrics~\cite{shi2022evaluation,roy2021reassessing,gros2020code}. As we are explicitly experimenting with two languages, allowing us to compare LLMs on both open-source and closed-source data, we could not use any readily available dataset. To ensure fairness and avoid duplicate data, we adopt the pipeline proposed by the CodeSearchNet dataset, and conduct cross-project evaluations, with lower risk of biases.  
To select open-source data (for all the tasks), we collected the top 10,000 starred projects for C\# and C++ with at least one commit after 2020. Table~\ref{tbl:dataset} presents the number of samples prepared for our experiments. It is important to note that LLMs (Language Model Models) are relatively slow and very costly. Therefore, following prior works and ensuring sufficient statistical strength for evaluation, we selected 1000 samples for code summarization in each setup. We discuss about the number of samples more in Section~\ref{datasets}. 

There have been ongoing debates regarding the choice of metrics for this task. CodeXGLUE benchmark recommends using BLEU-CN~\cite{lin2004orange}; a variant of BLEU~\cite{papineni2002bleu}, which is widely used in subsequent research. However, Shi \etal have proposed BLEU-DC as the metric that best reflects human perception~\cite{shi2022evaluation}. In our reporting and comparisons with other works, we primarily used BLEU-CN. Nonetheless, we also evaluated the performances using three other metrics, including BLEU-DC, ROUGE-L~\cite{lin2004rouge}, and METEOR~\cite{banerjee2005meteor}. For this study, we relied on some available metrics; our scripts can be adapted to other metrics.

\subsubsection{Code generation}
Code generation is the inverse of the code summarization task. The objective is to generate a function in a specific programming language (PL) based on a given natural language summary. We utilized the same dataset and approach as the code summarization task for this task but with swapped input and output. We evaluated the performance using two widely used and recommended metrics: BLEU-4, and CodeBLEU~\cite{ren2020codebleu}, which are part of the CodeXGLUE benchmark. We also applied CrystalBLEU~\cite{eghbali2022crystalbleu}, which was introduced after the CodeXGLUE Benchmark.

\subsection{Datasets}
\label{datasets}

\subsubsection{Open-source}

While datasets like CodeSearchNet are available for evaluating the tasks mentioned in Section ~\ref{tasks}, we needed to prepare new data for like-to-like comparison with our closed-source data. We note that C\# and C++ are the two primary languages used at \company. Following the approach used in collecting the CodeSearchNet dataset~\footnote{https://github.com/ncoop57/function\_parser}, we prepared data for our experiments by collecting the top 10,000 starred
projects for C\# and C++ with at least one commit after 2020. We collected code-comment pairs, performed de-duplication, and discarded samples with more than 400 tokens. We truncated the samples at 400 tokens to accommodate the context window limitations of the model. 
The final sample counts are presented in Table~\ref{tbl:dataset}.

\subsubsection{Closed-source}

For our closed source data-set we sampled source code files from the largest code bases across \company.  
Since some products at \company may include open source software or rely on open source libraries or toolchains, we manually ensured that the files included were not open source.
In addition, some products from \company have portions of their source code made available publicly via download or GitHub and we also manually checked to ensure that none of these projects were included in our sample. 
This provided certainty that none of the source files included in our closed-source sample had been included in the training of the models evaluated in this paper.  
Thus, to our knowledge, this is the first time that LLMs have been evaluated on code that we can be certain was not encountered during training.
\company software in the sample included developer tools, online services, desktop software, and internal infrastructure.
In an effort to ensure consistency with the Open-source dataset, we collected code-comment pairs, performed de-duplication, and discarded functions with more than 400 tokens. We collected 6,904 samples for C\# and 7,643 samples for C++ for code summarization and code generation tasks. For code completion tasks, similar to OSS projects, we collected 10,000 samples for both languages.

Note that it \emph{is} possible for a company to fine-tune pre-trained models on their own code bases in an effort to improve model performance.  This is both expensive and time-consuming
 and thus we posit is likely not undertaken by many companies.
Evaluating the impact of such fine-tuning is beyond the scope of this study as one of our goals is to understand how well these ``off-the-shelf'' models perform on unseen commercial codebases.

\subsection{Model(s) Used for Performance Evaluation}
In this paper, we utilized two different models, Code-Davinci-002 and GPT-3.5-Turbo, for the evaluation. We observed very similar findings with both models. Additionally, we employed another model, Text-embedding-ada-002, for embedding generation. Encoder-decoder models like CodeT5 were excluded from the evaluations because models such as Code-Davinci-002 have demonstrated state-of-the-art performance across various tasks~\cite{ahmed2022few,nashid2023retrieval}. Moreover, smaller encoder-decoder models require a significant amount of data for fine-tuning, which may not be readily available for many organizations.

\subsubsection{Code-Davinci-002}
The Codex~\cite{chen2021evaluating} models are part of the GPT-3~\cite{brown2020language} family, specifically designed and trained for code-related tasks. The training data for these models includes a large natural language corpus and billions of lines of publicly available code sourced from GitHub.
Our experiments use the Code-Davinci-002 model, a powerful GPT-3 variant leveraging 175 billion parameters. This model has been fine-tuned to excel at code-related tasks. Code-Davinci-002 is an upgraded and more advanced version of the original Codex model. It has undergone training on a more recent, large dataset, 
to enhance its capabilities and performance.

\subsubsection{GPT-3.5-Turbo}
GPT-3.5 models are highly proficient in comprehending and generating natural language and code. Among these models, Gpt-3.5-Turbo is the most capable and cost-effective option. Although it is primarily optimized for chat-based interactions, it also performs well in traditional completion tasks.

For our research, we use both these models: Code-Davinci-002, a completion model, and Gpt-3.5-Turbo, a chat model. Both models have unique strengths and will be utilized to assess the task's performance. Gpt-3.5-Turbo can generate until 4096 tokens, including the prompt tokens.

\subsection{Model(s) Used Embedding}
\subsubsection{Text-embedding-ada-002}
An embedding is a real-valued vector  trained to capture meaning in context, for natural language and code. It helps machine learning models and algorithms understand relationships within the content, enabling tasks like clustering or retrieval. Embeddings play a crucial role in applications such as knowledge retrieval and are used in various developer tools for retrieval augmented generation (RAG). We employ the text-embedding-ada-002 model in this paper. It's important to note that we solely utilized embeddings to illustrate the distinctions in open-source (OSS) and closed-source data in Section~\ref{embedding_dis}. 
This does not represent the actual performance of the language models on various software engineering tasks.

\begin{figure}[!t]
    \centering
    \includegraphics[width=0.95\textwidth]{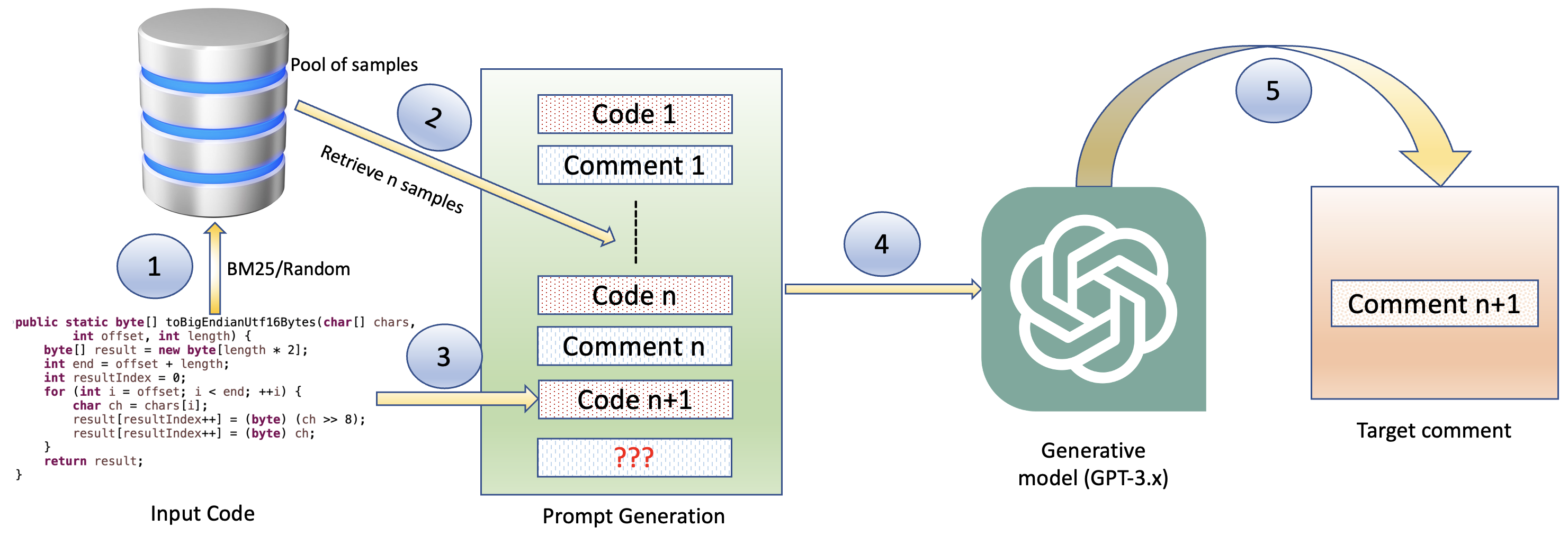}
    \caption{Different steps of our pipeline for code summarization task. (1) the target code is sent to the pool of samples, (2) n random/BM25 chosen samples are retrieved to build prompt, (3) prompt is build by appending the retrieved code-comment pair with the input code, (4) the prompt is sent to the GPT-3.x model, (5) target comment is extracted from the model.}
   \label{component}
\end{figure}


\section{Experiment: Methodology \& Results}

\subsection{RQ1: Performance Comparison}
\label{rq1}
\subsubsection{Methodology}
\label{rq1-method}

While we use a zero-shot setting for token and line completions,  for tasks like code summarization, we also used in-context learning. 
Autoregressive GPT models~\cite{radford2018improving,radford2019language,brown2020language,chen2021evaluating}, are primarily trained on the task of predicting the next token. 
For tasks like code summarization (or more extensive code generation) few-shot in-context learning appears to improve performance. 
In the following section, we will briefly discuss few-shot in-context learning,  and its variant that we apply for code summarization and code generation tasks.

\vspace{.2cm}

\noindent{\underline{\emph{In-Context Few-shot Learning. }}}
In-context learning refers to the approach of conditioning very large
language models (LLMs)  by providing instructions~\cite{ouyang2022training}, examples~\cite{wang2020generalizing}, or chain-of-thought~\cite{wei2022chain}
as text in a prompt. For LLMs, this approach is highly effective~\cite{brown2020language}, and offers
the advantage of not requiring costly parameter-updates. 
Few-shot  learning involves assembling a prompt with a few input-output pairs illustrating the operation of  a given task. For example, we can create a prompt with a few English sentences paired with their corresponding German translations when translating English to German. When we append a new English sentence to the prompt, the LLM can efficiently generate its German translation without requiring any weight updates. This technique is remarkably effective across various tasks.
Initially used in Natural Language Processing (NLP) applications, Few-shot  learning has demonstrated its effectiveness in a wide range of software engineering tasks. These tasks include code summarization~\cite{ahmed2022few}, code repair~\cite{nashid2023retrieval}, assertion generation~\cite{nashid2023retrieval}, code mutation~\cite{bareiss2022code}, test oracle generation from natural language documentation~\cite{bareiss2022code}, and test case generation~\cite{bareiss2022code}. Remarkably, Few-shot learning on these tasks has surpassed previous state-of-the-art models in several cases. We performed two kinds of few-shotting: 
first,  we randomly chose samples for each test instance from our training set; second, we retrieve particularly relevant examples using BM25, as described next.

\vspace{.2cm}

\noindent{\underline{\emph{Few-shot Learning with Retrieved Samples.}}}
BM25~\cite{robertson2009probabilistic} is a widely-used, high-performance ranking function for information retrieval and search engines, which substantially improves upon the traditional TF-IDF~\cite{ramos2003using} 
approach for 
assessing the relevance of documents to a given query. It achieves this by considering both the term frequency and document length while penalizing excessively long documents. 
Recently, Nashid \etal~\cite{nashid2023retrieval} utilized BM25 to select specific few-shot samples, significantly improving the performance of few-shot tasks for code repair and assertion generation. Similarly, Ahmed \etal~\cite{ahmed2023improving} found that this approach enhances code summarization performance. Nashid \etal found BM25 to be the more effective, even than neural methods, to find
relevant examples for few-shot learning. Figure~\ref{component} presents our pipeline for code summarization task. We followed a similar approach for code generation also. We must swap the code and NL descriptions to apply this same approach for the code generation task. 

For code summarization and code generation tasks, we utilized 4-shot learning due to the limitations imposed by the maximum token length of the model. \emph{To clarify, we exclusively utilized open-source code comment pairs as few-shot samples for the OSS experiment, while closed pairs were used as few-shot samples for the closed-source/proprietary experiment}.

\subsubsection{Result} 
We now present our findings regarding the models' performance on both open-source (OSS) and closed-source data across various tasks for two programming languages, C\# and C++

\vspace{.2cm}

\noindent{\underline{\emph{Token Completion.}}}
For both open-source (OSS) and closed-source data, we observed similar performance in C\# with both the Code-Davinci-002 and GPT-3.5-Turbo models. Table~\ref{tbl:token-com-res} illustrates that for C\#, in both open and closed-source data, the accuracy with the Code-Davinci-002 model is 71.32\% and 71.59\%, respectively, with a 10k sample in each category. We also conducted a proportioned Z-test and found no statistical significance.

On the other hand, in the case of C++, the performance decreased from 71.93\% to 64.42\% when transitioning from OSS to closed-source, with the Code-Davinci-002 model showing statistical significance. We obtained very similar results with the GPT-3.5-Turbo model. In summary, there is no significant difference in performance between OSS and closed-source C\# for token completion tasks. However, for C++, there is a significant drop in performance when transitioning from OSS to closed-source data. 

\vspace{.2cm}

\noindent{\underline{\emph{Line Completion.}}}
We observed similar results with line completion tasks as well. Unlike token completion, we now use two metrics: i) Exact match (EM) and ii) Edit similarity (ES).
Edit similarity (ES) uses a ratio function that calculates the standard Levenshtein distance~\cite{levenshtein1966binary} similarity ratio between two sequences, ranging from 0 to 100.
For EM, we applied the proportioned Z-test, and for ES, we employed an unpaired Wilcoxon signed-rank test. From Table~\ref{tbl:line-com-res}, we found once again that for C\#, there is no significant difference in performance with both metrics. However, for C++, the performance on closed-source significantly dropped compared to OSS. This trend is consistent across both models.



\vspace{.2cm}

\noindent{\underline{\emph{Code Summarization.}}}
Unlike token and line completions (zero-shot), we use a different setup for code summarization and code generation. We note that the main pre-training objectives for Language Models (LLMs) are next token prediction, enabling the model to perform well on token and line completions without any additional effort. However, this is not the case for code summarization. In code summarization, to enable the model to generate reasonable comments from code, we needed to use few-shot learning. We note that in real-world projects, comments often appear \emph{before} code, and so a model might struggle to autoregressively generate comments given code. Hence, few-shot learning becomes necessary in this context. We also tried an improved version of few-shot learning, using a retrieval algorithm to extract relevant samples from training/existing samples collection as few-shot samples rathen than randomly chosen ones.

Table~\ref{tbl:csum-res} shows the performance on both open-source (OSS) and closed-source datasets for C\# and C++. As before, for C\#, we observe no significant difference between OSS and closed-source datasets with the Code-Davinci-002 model. After applying BM25, the performance improved for both OSS and closed-source datasets, but the difference between them is statistically insignificant for all metrics used in our experiments. With the GPT-3.5-Turbo model, we observed some statistical significance, but the magnitude of the difference is low for all metrics.

For C++, when we utilized random samples instead of BM25-retrieved samples, we observed a slightly higher difference compared to what we observed with C\# using Code-Davinci-002. For one metric, BLEU-DC, the difference is statistically significant, but for all other metrics, we did not observe any significance. After applying BM25, the performance with OSS projects significantly improved, but for closed-source, the performance improves less. It's important to note that we have a much larger few-shot pool (142,195) for OSS than closed-source (6,643), which may have played a key role in the observed difference. With the GPT-3.5-Turbo model, we observed that with BM25, the magnitude of metrics increased for OSS projects, but we did not find any statistical significance due to the high variance of the data; GPT-3.5-Turbo in general is not as highly performant as Code-Davinic-002 at code-related tasks. 

In summary, for C\#, we did not find any difference for both few-shot learning and BM25-retrieved few-shot learning. However, for C++, although we haven't seen much difference with few-shot learning, with BM25, the difference increases, and it is statistically significant.

\vspace{.2cm}

\noindent{\underline{\emph{Code Generation.}}}
Unlike the other three tasks, for code generation, the models don't perform well even after applying few-shot learning (with or without BM25). While the models show promising results at the line level, they struggle to complete full functions. In the majority of cases with both models, we observe statistical significance, but the performance for both OSS and closed-source is on the lower side. A consistent pattern persists from the code summarization results. For C\#, with BM25, the magnitude of difference does not increase that much (less than 20\%), but for C++, the difference in magnitude increases by around 100\% (see Table~\ref{tbl:code-gen-res}).

\begin{table}[t]

\centering

\scriptsize
{%
\renewcommand{\arraystretch}{1.2}
\begin{tabular}{llccc}
\hline
\multicolumn{1}{c}{\multirow{2}{*}{Language}} & \multicolumn{1}{c}{\multirow{2}{*}{Model}} & \multicolumn{2}{c}{Category} & \multirow{2}{*}{p-value} \\ 
\multicolumn{1}{c}{}                          & \multicolumn{1}{c}{}                       & OSS      & Proprietary   &                          \\ \hline
\multirow{2}{*}{C\#}                          & Code-Davinci-002                           & 71.32    & 71.59         & 0.67                       \\
                                              & GPT-3.5-Turbo                              & 60.62       & 61.75           & 0.10                       \\
\cline{2-5}
\multirow{2}{*}{C++}                          & Code-Davinci-002                           & 71.93    & 64.42            & <0.01                       \\
                                              & GPT-3.5-Turbo                              & 62.42      & 58.38            & <0.01      \\ \hline                
\end{tabular}
}
\vspace{0.05in}
\caption{Models' Performance on OSS and Closed-source Token Completion (accuracy in \%)}
\vspace{-0.2in}
\label{tbl:token-com-res}
\end{table}

\begin{table}[t]

\centering

\scriptsize
{%
\renewcommand{\arraystretch}{1.2}
\begin{tabular}{llcccccc}
\hline
\multicolumn{1}{c}{\multirow{3}{*}{Language}} & \multicolumn{1}{c}{\multirow{3}{*}{Model}} & \multicolumn{3}{c}{EM}                              & \multicolumn{3}{c}{Edit Similarity}                 \\
\multicolumn{1}{c}{}                          & \multicolumn{1}{c}{}                       & \multicolumn{2}{c}{Category} & \multirow{2}{*}{p-value} & \multicolumn{2}{c}{Category} & \multirow{2}{*}{p-value} \\
\multicolumn{1}{c}{}                          & \multicolumn{1}{c}{}                       & OSS      & Proprietary   &                          & OSS      & Proprietary   &                          \\ \hline
\multirow{2}{*}{C\#}                          & Code-Davinci-002                           & 24.49    & 25.07            & 0.34                       & 60.94    & 59.17            & 0.65                       \\
                                              & GPT-3.5-Turbo                              & 17.25       & 16.75            & 0.35                       & 46.98       & 46.53            & 0.71                       \\
\cline{2-8}
\multirow{2}{*}{C++}                          & Code-Davinci-002                           & 33.46    & 21.63            & <0.01                       & 63.79    & 54.39            & <0.01                       \\
                                              & GPT-3.5-Turbo                              & 15.28       & 15.05            & 0.65                       & 48.17      & 44.66            & <0.01    \\ \hline                    
\end{tabular}
}
\vspace{0.05in}
\caption{Models' Performance on OSS and Closed-source Line Completion}
\vspace{-0.2in}
\label{tbl:line-com-res}
\end{table}

\begin{table}[t]
\scriptsize
\centering

\resizebox{\textwidth}{!}%
{%
\renewcommand{\arraystretch}{1.2}
\begin{tabular}{llc|ccc|ccc|ccc|ccc}

\hline

\multicolumn{1}{c}{\multirow{2}{*}{Language}} & \multicolumn{1}{c}{\multirow{2}{*}{Model}} & \multirow{2}{*}{\begin{tabular}[c]{@{}c@{}}BM25 \\ Applied?\end{tabular}} & \multicolumn{3}{c}{BLEU-CN} & \multicolumn{3}{c}{BLEU-DC} & \multicolumn{3}{c}{ROUGE-L} & \multicolumn{3}{c}{METEOR}  \\
\multicolumn{1}{c}{}                          & \multicolumn{1}{c}{}                       &                                                                           & OSS & Proprietary & p-value & OSS & Proprietary & p-value & OSS & Proprietary & p-value & OSS & Proprietary & p-value \\ \hline
\multirow{4}{*}{C\#}                          & \multirow{2}{*}{Code-Davinci-002}          & No                                                                        & 16.98 & 18.08         & 0.35     & 11.33 & 11.62         & 0.32     & 32.10 & 33.48         & 0.34    & 28.81 & 31.56         & 0.03     \\
                                              &                                            & Yes                                                                       & 21.15 & 21.96         & 0.03     & 15.75 & 16.08         & 0.07     & 35.37 & 37.03         & 0.11     & 32.85 & 35.01         & 0.02     \\
                                              \cline{2-15}
                                              & \multirow{2}{*}{GPT-3.5-Turbo}             & No                                                                        & 10.92 & 11.53         & 0.02     & 6.21 & 6.16         & 0.02     & 21.19 & 21.87         & 0.08     & 22.89 & 24.65        & <0.01     \\
                                              &                                            & Yes                                                                       & 13.73 & 14.48         & <0.01     & 8.52 & 8.77         & <0.01     & 21.86 & 23.50         & <0.01     & 23.11 & 25.65         & <0.01     \\
\hline
\multirow{4}{*}{C++}                          & \multirow{2}{*}{Code-Davinci-002}          & No                                                                        & 15.67 & 13.11         & 0.05     & 8.80 & 5.49         & <0.01     & 26.11 & 23.61         & 0.15     & 24.86 & 22.47         & 0.15     \\
                                              &                                            & Yes                                                                       & 26.64 & 16.74         & <0.01     & 20.48 & 9.21         & <0.01     & 35.37 & 26.59         & <0.01    & 34.43 & 25.59         & <0.01     \\
                                              \cline{2-15}
                                              & \multirow{2}{*}{GPT-3.5-Turbo}             & No                                                                        & 8.26 & 8.80         & 0.05     & 3.86 & 3.81         & 0.17     & 17.12 & 18.01         & 0.01     & 22.38 & 23.39         & 0.05     \\
                                              &                                            & Yes                                                                       & 14.46 & 10.69         & 0.24     & 8.94 & 5.24         & 0.26     & 22.11 & 18.81         & 0.51     & 26.38 & 23.52          & 0.65     \\ \hline
\end{tabular}
}
\vspace{0.05in}
\caption{Performance of GPT-3.x models on Code Summarization task with both open-source and closed-source projects. Note that we exclusively utilized
open-source code comment pairs as few-shot samples for the OSS experiment, while closed pairs were
used as few-shot samples for the closed-source/proprietary experiment.}
\vspace{-0.2in}
\label{tbl:csum-res}
\end{table}

\begin{table*}[t]
\scriptsize
\centering

\resizebox{.95\textwidth}{!}%
{%
\renewcommand{\arraystretch}{1.2}
\begin{tabular}{llc|ccc|ccc|ccc}
\hline
\multicolumn{1}{c}{\multirow{2}{*}{Language}} & \multicolumn{1}{c}{\multirow{2}{*}{Model}} & \multirow{2}{*}{\begin{tabular}[c]{@{}c@{}}BM25 \\ Applied?\end{tabular}} & \multicolumn{3}{c}{BLEU}      & \multicolumn{3}{c}{CodeBLEU}  & \multicolumn{3}{c}{CrystalBLEU} \\
\multicolumn{1}{c}{}                          & \multicolumn{1}{c}{}                       &                                                                           & OSS   & Proprietary & p-value & OSS   & Proprietary & p-value & OSS    & Proprietary  & p-value \\ \hline
\multirow{4}{*}{C\#}                          & \multirow{2}{*}{Code-Davinci-002}          & No                                                                        & 14.82 & 13.31       & <0.01     & 29.88 & 23.55       & <0.01     & 8.43   & 5.29          & 0.02     \\
                                              &                                            & Yes                                                                       & 15.41 & 14.79       & <0.01     & 30.23 & 24.84       & <0.01     & 10.04  & 7.22          & 0.01     \\ \cline{2-12}
                                              & \multirow{2}{*}{GPT-3.5-Turbo}             & No                                                                        & 10.02 & 7.72        & <0.01     & 23.23 & 21.07       & <0.01     & 6.24   & 5.50          & <0.01     \\
                                              &                                            & Yes                                                                       & 12.7  & 9.54        & <0.01     & 26.84 & 21.54      & <0.01     & 8.31   & 6.62          & <0.01     \\ \hline
\multirow{4}{*}{C++}                          & \multirow{2}{*}{Code-Davinci-002}          & No                                                                        & 7.53  & 5.14         & <0.01     & 16.9  & 13.27         & <0.01    & 3.45   & 1.59          & 0.19     \\
                                              &                                            & Yes                                                                       & 15.58 & 8.50         & <0.01     & 23.19 & 15.28         & <0.01    & 10.82  & 3.65          & <0.01     \\ \cline{2-12}
                                              & \multirow{2}{*}{GPT-3.5-Turbo}             & No                                                                        & 3.6   & 3.21         & <0.01     & 14.32 & 13.79         & <0.01     & 2.39   & 1.91          & 0.15     \\
                                              &                                            & Yes                                                                       & 7.6   & 3.76         & <0.01     & 18.65 & 12.56         & <0.01     & 8.15   & 2.29          & <0.01 \\ \hline   
\end{tabular}
}
\vspace{0.05in}
\caption{Performance of GPT-3.x models on Code Generation task with both open-source and closed-source projects. Note that we exclusively utilized
open-source code comment pairs as few-shot samples for the OSS experiment, while closed pairs were
used as few-shot samples for the closed-source/proprietary experiment.}
\vspace{-0.2in}
\label{tbl:code-gen-res}
\end{table*}

\subsection{RQ2: Improving Closed-Source Performance}
\label{rq2}
\subsubsection{Methodology}
Improving the model's performance with in-context learning on closed-source data can be challenging because there isn't enough data to enable few-shot learning (we encounter a very similar scenario). How can we enhance the model's performance on closed-source data? To address this question, we followed a similar approach as applied in Section \ref{rq1-method}. While employing BM25, instead of extracting relevant samples from a single specific source, we experimented with different sources (OSS, Closed, or both). Our findings indicate that the model's performance can be enhanced 
when few-shot learning involves samples from both sources rather than just one. This suggests that the model's performance on closed-source data can be improved by incorporating open-source data. We delve into the details of these results in the following subsection.


\subsubsection{Result} In this section, show that model performance on closed-source data can be improved by broadening the sources for few-shot sampling.

\vspace{.2cm}

\noindent{\underline{\emph{Code Summarization.}}}
In Section~\ref{rq1}, we did not combine or interchange the few-shot samples for both OSS and closed-source samples. Note that it can be a very common scenario where we have a lot more samples for few-shotting in OSS than in closed source (as seen in the case of C++). Can we use the samples from the OSS to improve performance? If the distribution or nature of closed-source samples aligns well with OSS samples, the models can benefit from having samples from the OSS. However, using only OSS samples may hurt the model performance because examples may not be relevant: indeed, intra-project or same-project few-shots usually provide better performance~\cite{ahmed2022few}. 

Therefore, we explored three sources for few-shot sampling using BM25 (OSS, closed-source, OSS+closed-source) in Table~\ref{csp}. We found that for C\#, pooling samples from OSS improves performance over pooling from closed-source samples. This may be because OSS and closed source are similar in nature, and using OSS increases the number of few-shot pool size, helping us find more relevant samples from the pool and thus improving performance. The best performance is achieved when combining OSS and closed-source for pooling samples with both Code-Davinci-002 and GPT-3.5-Turbo models for C\#. This indicates that the performance of the closed-source model can be improved using OSS data.

For C++, we observe something different 
than with C\#. 
When using OSS few-shots, the performance actually decreases relative 
to using closed-source few-shots, 
even though we have larger OSS sample pools (142,195). This indicates that C++ has a different nature, and it cannot benefit as much from OSS data, aligning with our findings in Section~\ref{rq1}. However, combining OSS and closed-source data gives the best performance, similar to what we observed with C\#.

\vspace{.2cm}

\noindent{\underline{\emph{Code Generation.}}}
Table~\ref{csc} presents the results for code generation. For C\#, we observed that the performance improves with OSS, and the best performance is achieved when we combine OSS and closed-source data for few-shot sampling, aligning exactly with the observation we made for C\# code summarization. However, for C++, similar to code summarization, the performance declines with OSS samples, and even combining OSS and closed-source does not yield the best performance. This further strengthens the claims we made about the uniqueness of closed-source C++.

\begin{table}[t]

\centering

\scriptsize
{%
\renewcommand{\arraystretch}{1.2}
\begin{tabular}{lllcccc}
\hline
\multicolumn{1}{c}{Language} & \multicolumn{1}{c}{Model}         & \multicolumn{1}{c}{Shot Source} & BLEU-CN & BLEU-DC & ROUGE-L & METEOR \\ \hline
\multirow{6}{*}{C\#}         & \multirow{3}{*}{Code-Davinci-002} & Open Source                     & 24.76   & 19.02   & 39.22   & 36.91  \\
                             &                                   & Closed Source                   & 21.96   & 16.08   & 37.03   & 35.01  \\ 
                             &                                   & Open+Closed Source               & \textbf{27.05}   & \textbf{21.27}   & \textbf{41.18}   & \textbf{39.15}  \\ \cline{2-7}
                             & \multirow{3}{*}{GPT-3.5-Turbo}    & Open Source                     & 15.35   & 9.31    & 23.92   & 25.78  \\
                             &                                   & Closed Source                   & 14.48   & 8.77    & 23.5    & 25.65  \\
                             &                                   & Open+Closed Source               & \textbf{16.95}   & \textbf{10.76}   & \textbf{25.12}   & \textbf{26.92}  \\ \hline
\multirow{6}{*}{C++}         & \multirow{3}{*}{Code-Davinci-002} & Open Source                     & 14.3    & 6.8     & 24.86   & 23.39  \\
                             &                                   & Closed Source                   & 16.74   & 9.21    & 26.59   & 25.59  \\
                             &                                   & Open+Closed Source               & \textbf{17.28}   & \textbf{9.96}    & \textbf{27.41}   & \textbf{25.98}  \\ \cline{2-7}
                             & \multirow{3}{*}{GPT-3.5-Turbo}    & Open Source                     & 9.13    & 3.92    & 17.7    & 22.44  \\
                             &                                   & Closed Source                   & 10.69   & 5.24    & 18.81   & 23.52  \\
                             &                                   & Open+Closed Source               & \textbf{10.81}   & \textbf{5.36}    & \textbf{18.92}   & \textbf{23.66} \\ \hline
\end{tabular}
}
\vspace{0.05in}
\caption{Models' performance on closed-source code summarization using different sources for few-shots. For C\#, 79.15\% and for C++, 67.25\% of the samples on average are retrieved from open-source, while we use both open- and closed-source as sources of shots.}
\vspace{-0.2in}
\label{csp}
\end{table}

\begin{table}[t]

\centering

\scriptsize
{%
\renewcommand{\arraystretch}{1.2}
\begin{tabular}{lllccc}
\hline
\multicolumn{1}{c}{Language} & \multicolumn{1}{c}{Model}         & \multicolumn{1}{c}{Shot Source} & BLEU  & Code BLEU & Crystal BLEU \\ \hline
\multirow{6}{*}{C\#}         & \multirow{3}{*}{Code-Davinci-002} & Open Source                     & 17.87 & 27.33     & 10.17        \\ 
                             &                                   & Closed Source                   & 14.79 & 24.84     & 7.22         \\
                             &                                   & Open+Closed Source               & \textbf{20.11} & \textbf{29.28}     & \textbf{11.92}        \\ \cline{2-6}
                             & \multirow{3}{*}{GPT-3.5-Turbo}    & Open Source                     & 9.73  & 22.92     & 8.54         \\
                             &                                   & Closed Source                   & 9.54  & 21.54     & 6.62         \\
                             &                                   & Open+Closed Source               & \textbf{10.18} & \textbf{23.47}     & \textbf{9.64}         \\ \hline
\multirow{6}{*}{C++}         & \multirow{3}{*}{Code-Davinci-002} & Open Source                     & 6.34  & 13.19     & 1.89         \\
                             &                                   & Closed Source                   & 8.5   & 15.28     & 3.65         \\
                             &                                   & Open+Closed Source               & \textbf{9.12}  & \textbf{15.77}     & \textbf{4.27}         \\ \cline{2-6}
                             & \multirow{3}{*}{GPT-3.5-Turbo}    & Open Source                     & 2.12  & 11.02     & 1.63         \\
                             &                                   & Closed Source                   & \textbf{3.76}  & \textbf{12.56}     & 2.29         \\
                             &                                   & Open+Closed Source               & 2.87  & 11.95     & \textbf{2.73}   \\ \hline
\end{tabular}
}
\vspace{0.05in}
\caption{Models' performance on closed-source code generation using different sources for few-shots. For C\#, 81.72\% and for C++, 87.40\% of the samples on average are retrieved from open-source, while we use both open- and closed-source as sources of shots.}
\vspace{-0.2in}
\label{csc}
\end{table}

\subsection{RQ3: Why are Models Failing on Closed-Source C++}

So far, we have found that the model performs worse with closed-source C++ data across all experimental setups than it does on open source C++.
We now examine this performance drop further. While our study cannot conclusively assert specific causal reasons, we can offer some helpful observations.

\subsubsection{Accuracy for Identifiers and Non-identifiers}

We began by comparing the model's performance on identifiers and non-identifiers. Every programming language has a fixed set of non-identifiers (operators,  keywords, delimiters, \emph{etc}); these tokens are used in very deterministic ways, and language models perform very well on these when handling token completion tasks
for non-identifers. The relative proportion of identifiers and non-identifiers are very similar for each language in both open-source (OSS) and closed-source contexts. 

Turning now to model performance: for C\#, there is no discernible difference of accuracy between OSS and closed-source identifiers and non-identifiers. However, in the case of C++, the accuracy of identifiers in closed-source dropped by 8-10\%, and for non-identifiers, it dropped by 4-6\% with different models (see Table~\ref{RQ4}). It is important to note that if the model fails to recognize an identifier, it can lead to poor performance on subsequent non-identifiers as well. The drop has been substantial, suggesting that it may be a contributing factor to the observed differences in performance.

\begin{table*}[t]
\scriptsize
\centering

\resizebox{.95\textwidth}{!}%
{%
\renewcommand{\arraystretch}{1.2}
\begin{tabular}{ll|cc|cc|cc|cc}
\hline
\multicolumn{1}{c}{\multirow{3}{*}{Langauge}} & \multicolumn{1}{c}{\multirow{3}{*}{Model}} & \multicolumn{4}{c}{Identifiers}                                                                      & \multicolumn{4}{c}{Non-identifiers}                                 \\
\multicolumn{1}{c}{}                          & \multicolumn{1}{c}{}                       & \multicolumn{2}{c}{Open-source}                     & \multicolumn{2}{c}{Closed-source}               & \multicolumn{2}{c}{Open-source}  & \multicolumn{2}{c}{Closed-source} \\
\multicolumn{1}{c}{}                          & \multicolumn{1}{c}{}                       & \# of samples         & Accuracy                    & \# of samples        & Accuracy                & \# of samples         & Accuracy & \# of samples         & Accuracy \\ \hline
\multirow{2}{*}{C\#}                          & Code-Davinci-002                           & \multirow{2}{*}{3662} & 53.65\%                     & \multirow{2}{*}{3562} & 53.31\%                     & \multirow{2}{*}{6338} & 81.52\%  & \multirow{2}{*}{6438}  & 81.70\%      \\
                                              & GPT-3.5-Turbo                              &                       & 44.15\%                     &                      & 45.82\%                    &                       & 70.13\%  &                       & 70.57\%      \\ \hline
\multirow{2}{*}{C++}                          & Code-Davinci-002                           & \multirow{2}{*}{3770} & 58.99\%                         & \multirow{2}{*}{3417} & 46.15\%                     & \multirow{2}{*}{6230} & 79.76\%      & \multirow{2}{*}{6583}  & 73.90\%      \\
                                              & GPT-3.5-Turbo                              &                       & 49.36\% &                      & 41.85\% &                       & 70.32\%  &                       & 66.96\%      \\ \hline
\end{tabular}
}
\vspace{0.05in}
\caption{Models' performance in generating identifiers and non-identifiers.}
\vspace{-0.2in}
\label{RQ4}
\end{table*}

\subsubsection{Observation on Length Distribution}
\begin{figure}[!t]
    \centering
    \includegraphics[scale=0.50]{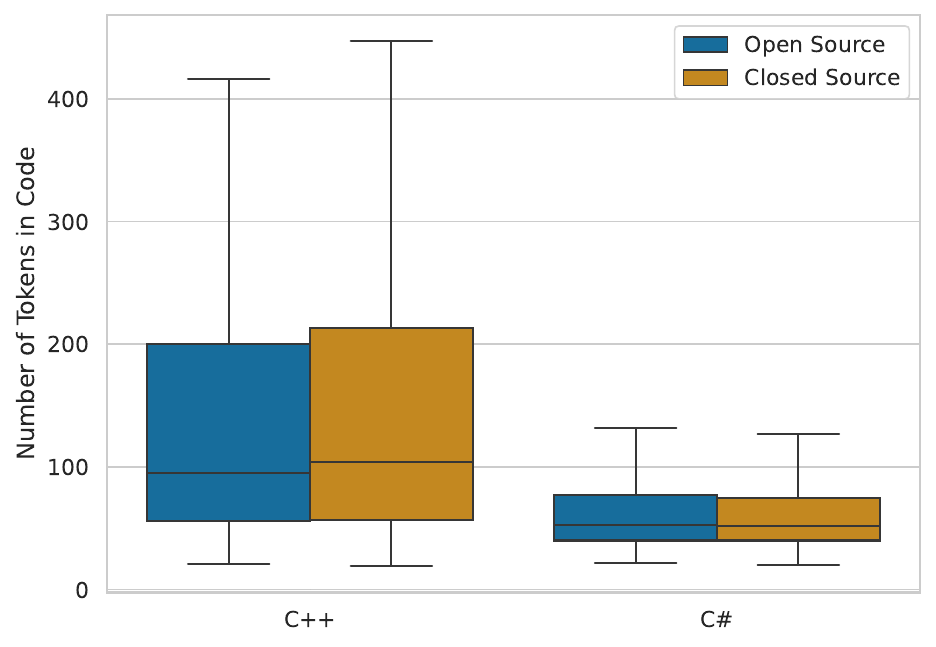}
    \caption{Boxplot presenting the lengths of OSS and closed-source functions.}
   \label{length_dis}
\end{figure}

To assess the impact of length distribution on the model's performance, we compared the length of open-source (OSS) and closed-source C\# and C++ functions\footnote{We discarded the functions with more than 400 tokens because of models' context limitations. Therefore, the findings we reported in this paper are strictly confined to the functions less than 400 tokens.}. We observed that there is no difference in length between the OSS and closed-source functions, both for C\# and C++ (see Figure~\ref{length_dis}). It is noteworthy that C\# functions are comparatively smaller than C++ functions. Additionally, we conducted a non-parametric unpaired Wilcoxon signed-rank test and found no statistically significant difference between OSS and closed-source length distributions. Consequently, the length does not have any impact on the model's performance.


\subsubsection{Observation on Sub-token Distribution}

Identifiers are crucial in the models' performance across various software engineering tasks~\cite{ahmed2022multilingual}. We investigated whether subtokens could elucidate the performance disparities in the models' execution on closed-source and open-source data. To achieve this, we extracted all the identifiers used in our samples (10,000 functions for each group) and segmented them into subtokens using the model tokenizer.

\begin{figure}[!t]
    \centering
    \includegraphics[scale=0.50] {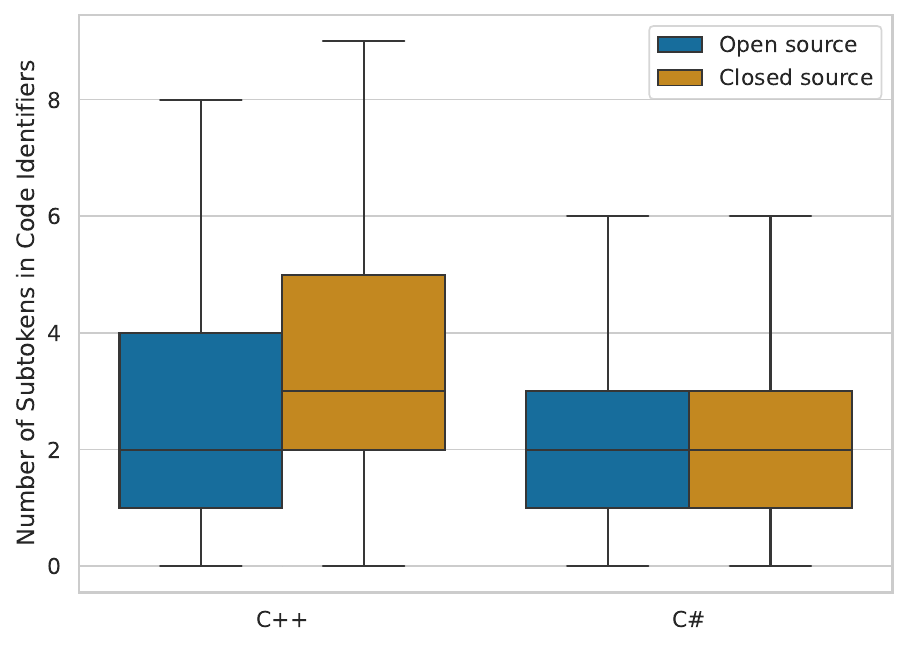}
    \caption{Sub-token count distribution of identifiers used in OSS and closed-source code.}
   \label{sub-token-dis}
\end{figure}

Our data indicates that for C\#, 
there is no discernible difference between open-source (OSS) and closed-source identifier subtoken count. However, for C++, OSS identifiers are longer than both C\# OSS and closed-source program identifiers, while closed-source C++ identifiers are longer than their OSS counterparts (refer to Figure~\ref{sub-token-dis}) and the difference is statistically significant. This observation also sheds light on the distinctive nature of code written in closed-source, indicating that identifiers in closed-source exhibit a different distribution compared to OSS data.

\subsubsection{Observation on BM25 Retrieved Samples}

\begin{figure}[!t]
    \centering
    \includegraphics[scale=0.35] {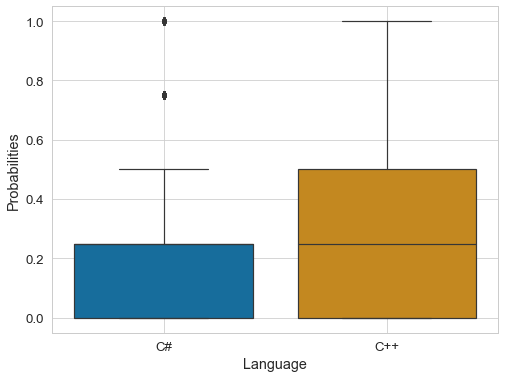}
    \caption{Probability of pooling samples from closed-source data while both OSS and closed-source functions are used as sample pools.}
   \label{BM25}
\end{figure}

In Section~\ref{rq2}, we demonstrated that closed-source performance can be enhanced by utilizing both open-source (OSS) and closed-source data as a sample pool. We also sought to comprehend the performance difference from the perspective of the BM25 algorithm. The BM25 algorithm aids in understanding the extent of similarity between OSS and closed-source functions. Figure~\ref{BM25} illustrates the probability of finding relevant samples from the closed-source data while choosing exemplars
(few-shots) for the code summarization task.

Upon comparing C\# and C++, it is evident that, for C++, we get a significantly higher fraction of samples from the closed source than C\#, while applying BM25 as a retrieval algorithm. Notably, although C++ has a much higher OSS + closed-source sample (142,195 + 6,643) compared to C\# (31,641 + 5,904), C++ still prefers relatively higher fraction of samples from the closed source (see Figure~\ref{BM25}) compared to C\# (32.75\% vs. 20.85\%). This indicates that closed-source C++ has a higher similarity with closed-sourced data. It's important to note that C++ closed-source may have a markedly different distribution from C++ open-source. Despite the larger OSS pool in C++, BM25 favors a higher amount of closed-sourced C++ samples compared to C\#.

\subsubsection{Observation on Embedding Space}
\label{embedding_dis}

We also examined the embedding space to study 
differences in closed and open-source C++ codes.
An embedding is a vector of floating-point numbers, that represents an input sequence of tokens, reflecting the patterns prevelant in the input text; 
the distance between two vectors measures the similarities of the texts. Small distances suggest high similarity, while large distances suggest low similarity. One caveat in our setting is that using embeddings directly from the model used for task-performance may not add any new information. To address this, we employed a separate model (text-embedding-ada-002)\footnote{\url{https://openai.com/blog/new-and-improved-embedding-model}} specifically designed for computing embeddings. 
We use the model to generate the embeddings for the code that we studied in this paper. 
We hypothesized that the embedding, also trained on the open-source software (OSS) data, should be able to reflect the performance differences we observed in our evaluation.

\begin{figure}[!t]
    \centering

    \begin{subfigure}[b]{0.45\textwidth}
        \centering
        \includegraphics[width=.85\linewidth]{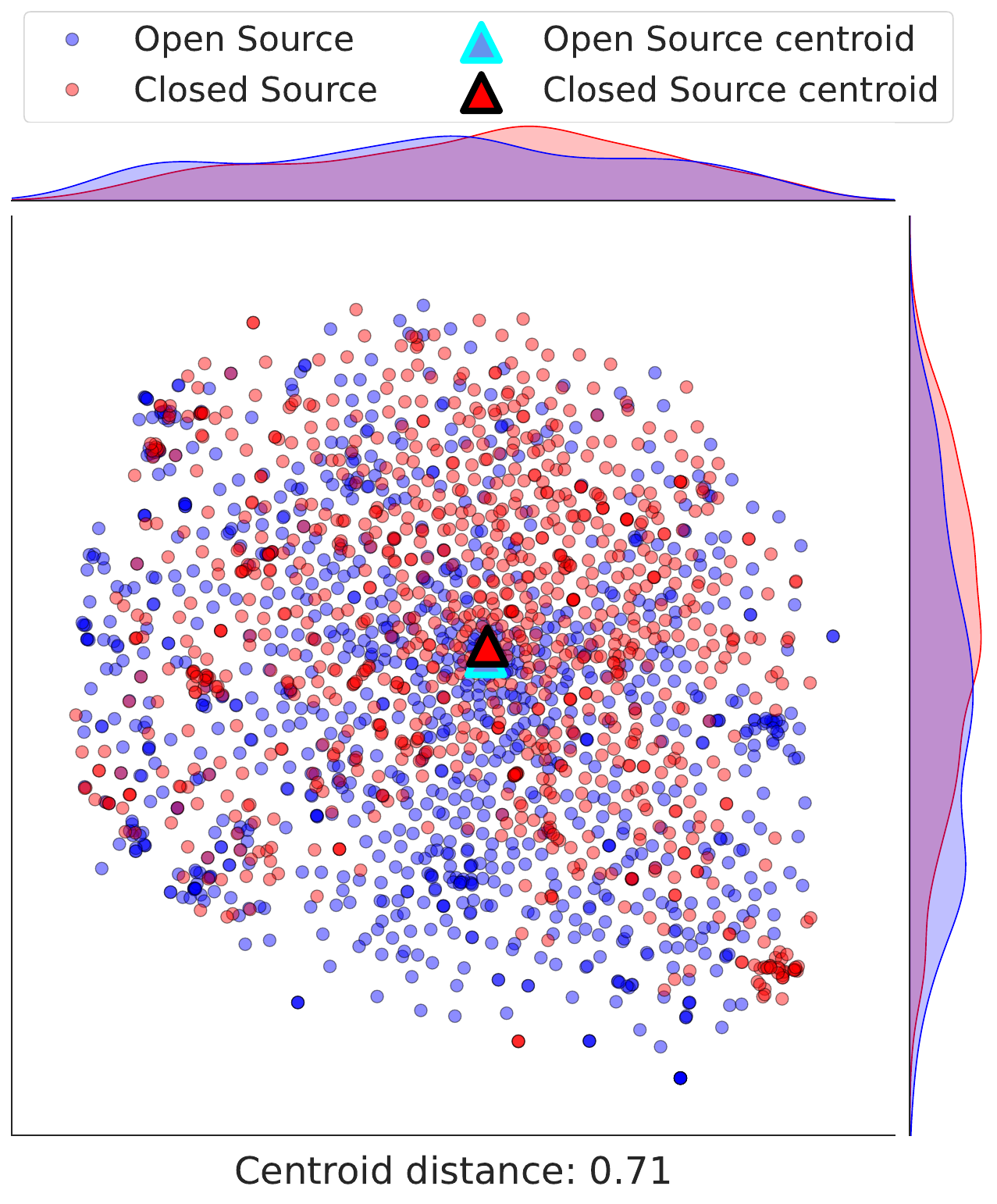}
        \caption{C\#}
        \label{sub:cs}
    \end{subfigure}%
    \begin{subfigure}[b]{0.45\textwidth}
        \centering
        \includegraphics[width=.85\linewidth]{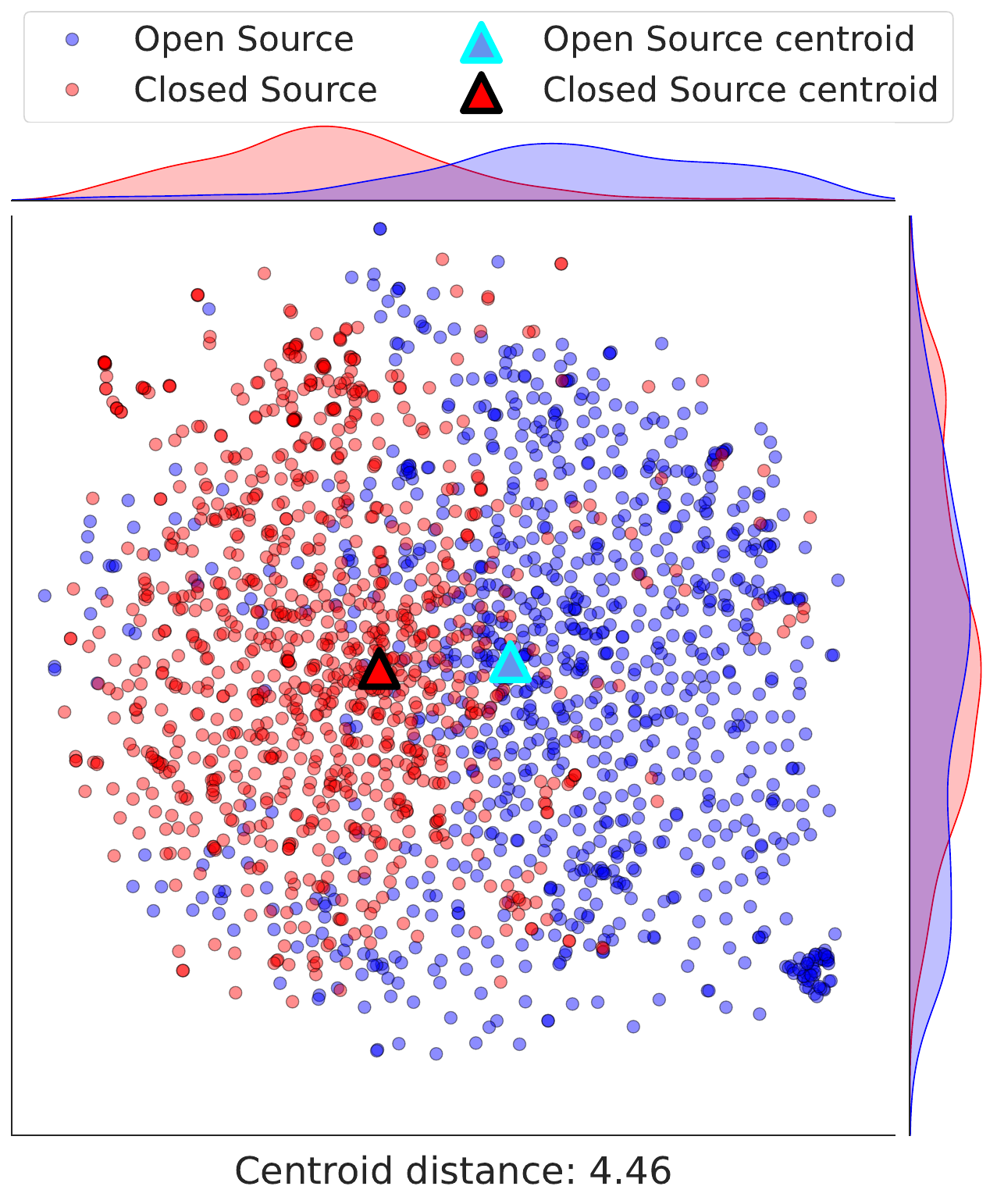} 
        \caption{C++}
        \label{sub:cpp}
    \end{subfigure}

    \caption{t-SNE plots showing the embedding distance between OSS and closed-sourced functions in two programming languages.}
    \label{fig:embedding}
\end{figure}

For visualizing the embeddings we use t-SNE~\cite{van2008visualizing} scattter plots, which gives us the opportunity to visually observe relationship between different code embeddings by projecting the embeddings in a lower dimensional space. Figure~\ref{fig:embedding} displays the t-SNE plots 
for both C\# and C++. We show 10,000 samples for each group, and these 10,000 samples are the same functions we employed in our token and line completion experiments. For C\#, the distribution of the samples overlaps, and the centroid distance between the open-source and closed-source is 0.71 (see Figure~\ref{sub:cs}). On the other hand, for C++, the centroid distance (4.46) is higher (see Figure~\ref{sub:cpp}), indicating that there are fewer similarities between functions in open-source (OSS) and closed-source C++ functions. The differing distributions of closed-source and open-source C++ suggest that the statistical patterns of usage of the language in the two corpora are different. 

\subsubsection{Key Takeaways}

We have observed various perspectives to explain the differences between open-source (OSS) and closed-source C++. We have determined that the length of the function does not impact performance. However, there are differences in identifier subtoken counts and the nature of the function, as reflected by the embedding. It is important to note that we do not know the reason for these differences; it could be organizational, related to project-specific practices, or influenced by the age of the code. We leave this aspect for future research. Nevertheless, we have demonstrated a method to detect when the model is likely to perform well or fail. If the distribution of samples from closed-source, as detected by the embedding, differs from that of OSS, it is likely that the model will perform poorly.

\section{Discussion \& Limitations}

\subsection{Experimentation with two programming languages}

In our study, the evaluation of LLMs' generalization performance is confined primarily to two programming languages: C\# and C++. 
This focus, while providing in-depth insights for these specific languages, might not generalize to others.
Different programming languages have unique syntax, idioms, and use-cases~\cite{allamanis2018survey,casalnuovo2019studying} 
which may limit generalizability.
This limitation is particularly relevant when considering languages with different paradigms (e.g., functional vs. object-oriented) or usage domains (e.g., web development vs. system programming). 
However, our choice to focus on C\# and C++ is still valid in certain contexts. 
Both languages are widely used in industry and academia, making our findings relevant to a significant portion of the software development community~\cite{meyerovich2013empirical}. 
Additionally, C\# and C++ offer a balance between modern, high-level programming features and low-level control, making them representative of a broad range of programming challenges and scenarios. 
Therefore, while our study's language scope is a limitation, it arguably still provides insights into potential Open/Closed source performance differences. 

\subsection{Sample Sizes}
\label{sample}

For code completions, we have 10,000 samples from each OSS and closed-source project, which are sufficient for achieving statistical significance, based on summary statistics used. However, for code summarization and code generation, we have far fewer samples (1k). 
To verify experimental power, we used the observed means and standard deviations to calculate the required sample sizes (using G*power~\cite{faul2009statistical,faul2007g}), employing commonly used values: a desired p-value ($\alpha$) of 0.01 and a power (1-$\beta$) of 0.80. Using the Wilcoxon Signed-rank test, we found that the needed sample size was consistently below the sample size 1000. 




\subsection{No access to the training data}
Lack of access to the models' training data significantly limits our ability to investigate the models' generalization and memorization capabilities.
The question of how much of what a model generates is the result of memorized code snippets and how much is the result of generalization is a key question in evaluating the utility of such models. Though we can not directly perform such evaluation, we can ensure that the closed-source data used for performance evaluation were never part of training data of these model. Performing well on closed-source data gives us some confidence on the models' generalization capability.

\section{Related Work}

\subsection{LLMs in SE} LLMs are widely used in software engineering tasks, including code summarization~\cite{chen2021evaluating,ahmed2022few,ahmed2023improving}, code generation~\cite{chen2021evaluating,xu2022systematic,nijkamp2022codegen,nijkamp2023codegen2, roziere2023code}, program repair~\cite{wei2023copiloting, nashid2023retrieval, ahmed2023better,jiang2023impact,fan2023automated}, and testing~\cite{kang2023large, feldt2023towards}. Several new ideas are evolving, such as chain-of-thoughts~\cite{wei2022chain}, self-consistency~\cite{wang2022self}, and retrieval augmented generation (RAG)~\cite{nashid2023retrieval}. Few-shot learning has mostly been used for in-context learning, and other ideas are still under consideration. However, no study has been performed comparing OSS and closed-source data in different OSS datasets. Studying OSS and closed-source data addresses two questions: i) Can LLMs perform as well on closed-source samples not seen in the training data? ii) Can OSS data help (\emph{via} in-context learning) improve performance on closed-source data? In this paper, we have tried to answer these two questions. Though we could not precisely justify the reasons for differences, we provided some possible directions for future studies.

\subsection{LLMs Generalization}

In prior studies with deep learning models, it was found that the model's performance on open-source and proprietary samples was comparable~\cite{pradel2020typewriter}. However, these earlier models had relatively smaller training datasets that were easily accessible. With the emergence of Large Language Models (LLMs), which are trained on billions of tokens, it becomes challenging to assess their generalization ability. Existing datasets such as HumanEval~\cite{chen2021evaluating}, HUMANEVAL  +~\cite{liu2023your}, and CoderEval~\cite{yu2023codereval} are used to evaluate the model's code synthesis capability. Still, they are often limited in size (a few hundred samples) and primarily consist of self-contained algorithmic problems, which may not represent real-world software development scenarios. Establishing a connection between open-source and proprietary projects is crucial to building trust with industry users.
Rabin et al. conducted experiments to measure the extent of memorization in models by introducing random noise to the original training dataset~\cite{rabin2023memorization}. They used various metrics to quantify the impact of noise on different aspects of training and testing. The findings indicated that all models showed some degree of memorization. This could be concerning for code intelligence tasks that rely on noisy and repetitive data sources, such as code from GitHub. However, it's important to note that the models evaluated in their study (Code2Seq~\cite{alon2018code2seq}, Code2Vec~\cite{alon2019code2vec}, GGNN~\cite{allamanis2018learning}, GREAT~\cite{hellendoorn2019global}, Transformers~\cite{vaswani2017attention}, CodeBERT~\cite{feng2020codebert}) significantly differ from the recent LLMs' size and nature. Moreover, conducting such studies without access to the LLMs' training data is not feasible.

\section{Conclusion}
Our study investigated the performance of two OpenAI models (Code-Davinci-002 and GPT-3.5-Turbo) on publicly available open-source and non-public, proprietary codes to observe the differences between them regarding the models' performance on four popular SE tasks. We found minimal differences in code completion (both token and line level) performance for C\#, but statistically significant differences for C++. Similarly, we observed fewer discrepancies in few-shot performance for code summarization and generation in C\# compared to C++. While the causal reasons behind these differences remain inconclusive, our findings suggest potential directions for further research. Additionally, we demonstrated that leveraging few-shot samples from both open-source and closed-source code can enhance model performance on closed-source data.

\vspace{.3cm}

\noindent{\emph{Acknowledgement.}}
Ahmed and Devanbu are partially funded by National Science
Foundation under Grant NSF CCF (SHF-MEDIUM) No. 2107592 and
the Intelligence Advanced Research Projects Agency (IARPA) under
contract W911NF20C0038. Our conclusions do not necessarily reflect the position or the policy of the
sponsors and no official endorsement should be inferred.

\bibliographystyle{ACM-Reference-Format}
\bibliography{reference}


\begin{thebibliography}{81}


\ifx \showCODEN    \undefined \def \showCODEN     #1{\unskip}     \fi
\ifx \showDOI      \undefined \def \showDOI       #1{#1}\fi
\ifx \showISBNx    \undefined \def \showISBNx     #1{\unskip}     \fi
\ifx \showISBNxiii \undefined \def \showISBNxiii  #1{\unskip}     \fi
\ifx \showISSN     \undefined \def \showISSN      #1{\unskip}     \fi
\ifx \showLCCN     \undefined \def \showLCCN      #1{\unskip}     \fi
\ifx \shownote     \undefined \def \shownote      #1{#1}          \fi
\ifx \showarticletitle \undefined \def \showarticletitle #1{#1}   \fi
\ifx \showURL      \undefined \def \showURL       {\relax}        \fi
\providecommand\bibfield[2]{#2}
\providecommand\bibinfo[2]{#2}
\providecommand\natexlab[1]{#1}
\providecommand\showeprint[2][]{arXiv:#2}

\bibitem[Ahmad et~al\mbox{.}(2020)]%
        {ahmad2020transformer}
\bibfield{author}{\bibinfo{person}{Wasi Ahmad}, \bibinfo{person}{Saikat
  Chakraborty}, \bibinfo{person}{Baishakhi Ray}, {and} \bibinfo{person}{Kai-Wei
  Chang}.} \bibinfo{year}{2020}\natexlab{}.
\newblock \showarticletitle{A Transformer-based Approach for Source Code
  Summarization}. In \bibinfo{booktitle}{\emph{Proceedings of the 58th Annual
  Meeting of the Association for Computational Linguistics}}.
  \bibinfo{pages}{4998--5007}.
\newblock


\bibitem[Ahmed and Devanbu(2022a)]%
        {ahmed2022few}
\bibfield{author}{\bibinfo{person}{Toufique Ahmed} {and}
  \bibinfo{person}{Premkumar Devanbu}.} \bibinfo{year}{2022}\natexlab{a}.
\newblock \showarticletitle{Few-shot training LLMs for project-specific
  code-summarization}. In \bibinfo{booktitle}{\emph{37th IEEE/ACM International
  Conference on Automated Software Engineering}}. \bibinfo{pages}{1--5}.
\newblock


\bibitem[Ahmed and Devanbu(2022b)]%
        {ahmed2022multilingual}
\bibfield{author}{\bibinfo{person}{Toufique Ahmed} {and}
  \bibinfo{person}{Premkumar Devanbu}.} \bibinfo{year}{2022}\natexlab{b}.
\newblock \showarticletitle{Multilingual training for software engineering}. In
  \bibinfo{booktitle}{\emph{Proceedings of the 44th International Conference on
  Software Engineering}}. \bibinfo{pages}{1443--1455}.
\newblock


\bibitem[Ahmed and Devanbu(2023)]%
        {ahmed2023better}
\bibfield{author}{\bibinfo{person}{Toufique Ahmed} {and}
  \bibinfo{person}{Premkumar Devanbu}.} \bibinfo{year}{2023}\natexlab{}.
\newblock \showarticletitle{Better patching using LLM prompting, via
  Self-Consistency}. In \bibinfo{booktitle}{\emph{2023 38th IEEE/ACM
  International Conference on Automated Software Engineering (ASE)}}. IEEE,
  \bibinfo{pages}{1742--1746}.
\newblock


\bibitem[Ahmed et~al\mbox{.}(2024)]%
        {ahmed2023improving}
\bibfield{author}{\bibinfo{person}{Toufique Ahmed},
  \bibinfo{person}{Kunal~Suresh Pai}, \bibinfo{person}{Premkumar Devanbu},
  {and} \bibinfo{person}{Earl~T Barr}.} \bibinfo{year}{2024}\natexlab{}.
\newblock \showarticletitle{Automatic Semantic Augmentation of Language Model
  Prompts (for Code Summarization)}.
\newblock \bibinfo{journal}{\emph{ICSE}} (\bibinfo{year}{2024}).
\newblock


\bibitem[Allamanis et~al\mbox{.}(2018a)]%
        {allamanis2018survey}
\bibfield{author}{\bibinfo{person}{Miltiadis Allamanis},
  \bibinfo{person}{Earl~T Barr}, \bibinfo{person}{Premkumar Devanbu}, {and}
  \bibinfo{person}{Charles Sutton}.} \bibinfo{year}{2018}\natexlab{a}.
\newblock \showarticletitle{A survey of machine learning for big code and
  naturalness}.
\newblock \bibinfo{journal}{\emph{ACM Computing Surveys (CSUR)}}
  \bibinfo{volume}{51}, \bibinfo{number}{4} (\bibinfo{year}{2018}),
  \bibinfo{pages}{1--37}.
\newblock


\bibitem[Allamanis et~al\mbox{.}(2018b)]%
        {allamanis2018learning}
\bibfield{author}{\bibinfo{person}{Miltiadis Allamanis}, \bibinfo{person}{Marc
  Brockschmidt}, {and} \bibinfo{person}{Mahmoud Khademi}.}
  \bibinfo{year}{2018}\natexlab{b}.
\newblock \showarticletitle{Learning to Represent Programs with Graphs}. In
  \bibinfo{booktitle}{\emph{International Conference on Learning
  Representations}}.
\newblock


\bibitem[Alon et~al\mbox{.}(2018)]%
        {alon2018code2seq}
\bibfield{author}{\bibinfo{person}{Uri Alon}, \bibinfo{person}{Shaked Brody},
  \bibinfo{person}{Omer Levy}, {and} \bibinfo{person}{Eran Yahav}.}
  \bibinfo{year}{2018}\natexlab{}.
\newblock \showarticletitle{code2seq: Generating sequences from structured
  representations of code}.
\newblock \bibinfo{journal}{\emph{arXiv preprint arXiv:1808.01400}}
  (\bibinfo{year}{2018}).
\newblock


\bibitem[Alon et~al\mbox{.}(2019)]%
        {alon2019code2vec}
\bibfield{author}{\bibinfo{person}{Uri Alon}, \bibinfo{person}{Meital
  Zilberstein}, \bibinfo{person}{Omer Levy}, {and} \bibinfo{person}{Eran
  Yahav}.} \bibinfo{year}{2019}\natexlab{}.
\newblock \showarticletitle{code2vec: Learning distributed representations of
  code}.
\newblock \bibinfo{journal}{\emph{Proceedings of the ACM on Programming
  Languages}} \bibinfo{volume}{3}, \bibinfo{number}{POPL}
  (\bibinfo{year}{2019}), \bibinfo{pages}{1--29}.
\newblock


\bibitem[Banerjee and Lavie(2005)]%
        {banerjee2005meteor}
\bibfield{author}{\bibinfo{person}{Satanjeev Banerjee} {and}
  \bibinfo{person}{Alon Lavie}.} \bibinfo{year}{2005}\natexlab{}.
\newblock \showarticletitle{METEOR: An automatic metric for MT evaluation with
  improved correlation with human judgments}. In
  \bibinfo{booktitle}{\emph{Proceedings of the acl workshop on intrinsic and
  extrinsic evaluation measures for machine translation and/or summarization}}.
  \bibinfo{pages}{65--72}.
\newblock


\bibitem[Barei{\ss} et~al\mbox{.}(2022)]%
        {bareiss2022code}
\bibfield{author}{\bibinfo{person}{Patrick Barei{\ss}},
  \bibinfo{person}{Beatriz Souza}, \bibinfo{person}{Marcelo d'Amorim}, {and}
  \bibinfo{person}{Michael Pradel}.} \bibinfo{year}{2022}\natexlab{}.
\newblock \showarticletitle{Code generation tools (almost) for free? a study of
  few-shot, pre-trained language models on code}.
\newblock \bibinfo{journal}{\emph{arXiv preprint arXiv:2206.01335}}
  (\bibinfo{year}{2022}).
\newblock


\bibitem[Bird et~al\mbox{.}(2022)]%
        {bird2022taking}
\bibfield{author}{\bibinfo{person}{Christian Bird}, \bibinfo{person}{Denae
  Ford}, \bibinfo{person}{Thomas Zimmermann}, \bibinfo{person}{Nicole
  Forsgren}, \bibinfo{person}{Eirini Kalliamvakou}, \bibinfo{person}{Travis
  Lowdermilk}, {and} \bibinfo{person}{Idan Gazit}.}
  \bibinfo{year}{2022}\natexlab{}.
\newblock \showarticletitle{Taking Flight with Copilot: Early insights and
  opportunities of AI-powered pair-programming tools}.
\newblock \bibinfo{journal}{\emph{Queue}} \bibinfo{volume}{20},
  \bibinfo{number}{6} (\bibinfo{year}{2022}), \bibinfo{pages}{35--57}.
\newblock


\bibitem[Brown et~al\mbox{.}(2020)]%
        {brown2020language}
\bibfield{author}{\bibinfo{person}{Tom Brown}, \bibinfo{person}{Benjamin Mann},
  \bibinfo{person}{Nick Ryder}, \bibinfo{person}{Melanie Subbiah},
  \bibinfo{person}{Jared~D Kaplan}, \bibinfo{person}{Prafulla Dhariwal},
  \bibinfo{person}{Arvind Neelakantan}, \bibinfo{person}{Pranav Shyam},
  \bibinfo{person}{Girish Sastry}, \bibinfo{person}{Amanda Askell},
  {et~al\mbox{.}}} \bibinfo{year}{2020}\natexlab{}.
\newblock \showarticletitle{Language models are few-shot learners}.
\newblock \bibinfo{journal}{\emph{Advances in neural information processing
  systems}}  \bibinfo{volume}{33} (\bibinfo{year}{2020}),
  \bibinfo{pages}{1877--1901}.
\newblock


\bibitem[Casalnuovo et~al\mbox{.}(2019)]%
        {casalnuovo2019studying}
\bibfield{author}{\bibinfo{person}{Casey Casalnuovo}, \bibinfo{person}{Kenji
  Sagae}, {and} \bibinfo{person}{Prem Devanbu}.}
  \bibinfo{year}{2019}\natexlab{}.
\newblock \showarticletitle{Studying the difference between natural and
  programming language corpora}.
\newblock \bibinfo{journal}{\emph{Empirical Software Engineering}}
  \bibinfo{volume}{24} (\bibinfo{year}{2019}), \bibinfo{pages}{1823--1868}.
\newblock


\bibitem[Chen et~al\mbox{.}(2021)]%
        {chen2021evaluating}
\bibfield{author}{\bibinfo{person}{Mark Chen}, \bibinfo{person}{Jerry Tworek},
  \bibinfo{person}{Heewoo Jun}, \bibinfo{person}{Qiming Yuan},
  \bibinfo{person}{Henrique Ponde de~Oliveira Pinto}, \bibinfo{person}{Jared
  Kaplan}, \bibinfo{person}{Harri Edwards}, \bibinfo{person}{Yuri Burda},
  \bibinfo{person}{Nicholas Joseph}, \bibinfo{person}{Greg Brockman},
  {et~al\mbox{.}}} \bibinfo{year}{2021}\natexlab{}.
\newblock \showarticletitle{Evaluating large language models trained on code}.
\newblock \bibinfo{journal}{\emph{arXiv preprint arXiv:2107.03374}}
  (\bibinfo{year}{2021}).
\newblock


\bibitem[Dong et~al\mbox{.}(2022)]%
        {dong2022survey}
\bibfield{author}{\bibinfo{person}{Qingxiu Dong}, \bibinfo{person}{Lei Li},
  \bibinfo{person}{Damai Dai}, \bibinfo{person}{Ce Zheng},
  \bibinfo{person}{Zhiyong Wu}, \bibinfo{person}{Baobao Chang},
  \bibinfo{person}{Xu Sun}, \bibinfo{person}{Jingjing Xu}, {and}
  \bibinfo{person}{Zhifang Sui}.} \bibinfo{year}{2022}\natexlab{}.
\newblock \showarticletitle{A survey for in-context learning}.
\newblock \bibinfo{journal}{\emph{arXiv preprint arXiv:2301.00234}}
  (\bibinfo{year}{2022}).
\newblock


\bibitem[Eddy et~al\mbox{.}(2013)]%
        {eddy2013evaluating}
\bibfield{author}{\bibinfo{person}{Brian~P Eddy}, \bibinfo{person}{Jeffrey~A
  Robinson}, \bibinfo{person}{Nicholas~A Kraft}, {and}
  \bibinfo{person}{Jeffrey~C Carver}.} \bibinfo{year}{2013}\natexlab{}.
\newblock \showarticletitle{Evaluating source code summarization techniques:
  Replication and expansion}. In \bibinfo{booktitle}{\emph{2013 21st
  International Conference on Program Comprehension (ICPC)}}. IEEE,
  \bibinfo{pages}{13--22}.
\newblock


\bibitem[Eghbali and Pradel(2022)]%
        {eghbali2022crystalbleu}
\bibfield{author}{\bibinfo{person}{Aryaz Eghbali} {and}
  \bibinfo{person}{Michael Pradel}.} \bibinfo{year}{2022}\natexlab{}.
\newblock \showarticletitle{CrystalBLEU: precisely and efficiently measuring
  the similarity of code}. In \bibinfo{booktitle}{\emph{Proceedings of the 37th
  IEEE/ACM International Conference on Automated Software Engineering}}.
  \bibinfo{pages}{1--12}.
\newblock


\bibitem[Fan et~al\mbox{.}(2023b)]%
        {fan2023large}
\bibfield{author}{\bibinfo{person}{Angela Fan}, \bibinfo{person}{Beliz
  Gokkaya}, \bibinfo{person}{Mark Harman}, \bibinfo{person}{Mitya Lyubarskiy},
  \bibinfo{person}{Shubho Sengupta}, \bibinfo{person}{Shin Yoo}, {and}
  \bibinfo{person}{Jie~M Zhang}.} \bibinfo{year}{2023}\natexlab{b}.
\newblock \showarticletitle{Large language models for software engineering:
  Survey and open problems}.
\newblock \bibinfo{journal}{\emph{arXiv preprint arXiv:2310.03533}}
  (\bibinfo{year}{2023}).
\newblock


\bibitem[Fan et~al\mbox{.}(2023a)]%
        {fan2023automated}
\bibfield{author}{\bibinfo{person}{Zhiyu Fan}, \bibinfo{person}{Xiang Gao},
  \bibinfo{person}{Martin Mirchev}, \bibinfo{person}{Abhik Roychoudhury}, {and}
  \bibinfo{person}{Shin~Hwei Tan}.} \bibinfo{year}{2023}\natexlab{a}.
\newblock \showarticletitle{Automated repair of programs from large language
  models}. In \bibinfo{booktitle}{\emph{2023 IEEE/ACM 45th International
  Conference on Software Engineering (ICSE)}}. IEEE,
  \bibinfo{pages}{1469--1481}.
\newblock


\bibitem[Faul et~al\mbox{.}(2009)]%
        {faul2009statistical}
\bibfield{author}{\bibinfo{person}{Franz Faul}, \bibinfo{person}{Edgar
  Erdfelder}, \bibinfo{person}{Axel Buchner}, {and}
  \bibinfo{person}{Albert-Georg Lang}.} \bibinfo{year}{2009}\natexlab{}.
\newblock \showarticletitle{Statistical power analyses using G* Power 3.1:
  Tests for correlation and regression analyses}.
\newblock \bibinfo{journal}{\emph{Behavior research methods}}
  \bibinfo{volume}{41}, \bibinfo{number}{4} (\bibinfo{year}{2009}),
  \bibinfo{pages}{1149--1160}.
\newblock


\bibitem[Faul et~al\mbox{.}(2007)]%
        {faul2007g}
\bibfield{author}{\bibinfo{person}{Franz Faul}, \bibinfo{person}{Edgar
  Erdfelder}, \bibinfo{person}{Albert-Georg Lang}, {and} \bibinfo{person}{Axel
  Buchner}.} \bibinfo{year}{2007}\natexlab{}.
\newblock \showarticletitle{G* Power 3: A flexible statistical power analysis
  program for the social, behavioral, and biomedical sciences}.
\newblock \bibinfo{journal}{\emph{Behavior research methods}}
  \bibinfo{volume}{39}, \bibinfo{number}{2} (\bibinfo{year}{2007}),
  \bibinfo{pages}{175--191}.
\newblock


\bibitem[Feldt et~al\mbox{.}(2023)]%
        {feldt2023towards}
\bibfield{author}{\bibinfo{person}{Robert Feldt}, \bibinfo{person}{Sungmin
  Kang}, \bibinfo{person}{Juyeon Yoon}, {and} \bibinfo{person}{Shin Yoo}.}
  \bibinfo{year}{2023}\natexlab{}.
\newblock \showarticletitle{Towards Autonomous Testing Agents via
  Conversational Large Language Models}.
\newblock \bibinfo{journal}{\emph{arXiv preprint arXiv:2306.05152}}
  (\bibinfo{year}{2023}).
\newblock


\bibitem[Feng et~al\mbox{.}(2020)]%
        {feng2020codebert}
\bibfield{author}{\bibinfo{person}{Zhangyin Feng}, \bibinfo{person}{Daya Guo},
  \bibinfo{person}{Duyu Tang}, \bibinfo{person}{Nan Duan},
  \bibinfo{person}{Xiaocheng Feng}, \bibinfo{person}{Ming Gong},
  \bibinfo{person}{Linjun Shou}, \bibinfo{person}{Bing Qin},
  \bibinfo{person}{Ting Liu}, \bibinfo{person}{Daxin Jiang}, {et~al\mbox{.}}}
  \bibinfo{year}{2020}\natexlab{}.
\newblock \showarticletitle{CodeBERT: A Pre-Trained Model for Programming and
  Natural Languages}. In \bibinfo{booktitle}{\emph{Proceedings of the 2020
  Conference on Empirical Methods in Natural Language Processing: Findings}}.
  \bibinfo{pages}{1536--1547}.
\newblock


\bibitem[Gros et~al\mbox{.}(2020)]%
        {gros2020code}
\bibfield{author}{\bibinfo{person}{David Gros}, \bibinfo{person}{Hariharan
  Sezhiyan}, \bibinfo{person}{Prem Devanbu}, {and} \bibinfo{person}{Zhou Yu}.}
  \bibinfo{year}{2020}\natexlab{}.
\newblock \showarticletitle{Code to Comment ?Translation?: Data, Metrics,
  Baselining \& Evaluation}. In \bibinfo{booktitle}{\emph{2020 35th IEEE/ACM
  International Conference on Automated Software Engineering (ASE)}}. IEEE,
  \bibinfo{pages}{746--757}.
\newblock


\bibitem[Haiduc et~al\mbox{.}(2010a)]%
        {haiduc2010supporting}
\bibfield{author}{\bibinfo{person}{Sonia Haiduc}, \bibinfo{person}{Jairo
  Aponte}, {and} \bibinfo{person}{Andrian Marcus}.}
  \bibinfo{year}{2010}\natexlab{a}.
\newblock \showarticletitle{Supporting program comprehension with source code
  summarization}. In \bibinfo{booktitle}{\emph{Proceedings of the 32nd ACM/IEEE
  International Conference on Software Engineering-Volume 2}}.
  \bibinfo{pages}{223--226}.
\newblock


\bibitem[Haiduc et~al\mbox{.}(2010b)]%
        {haiduc2010use}
\bibfield{author}{\bibinfo{person}{Sonia Haiduc}, \bibinfo{person}{Jairo
  Aponte}, \bibinfo{person}{Laura Moreno}, {and} \bibinfo{person}{Andrian
  Marcus}.} \bibinfo{year}{2010}\natexlab{b}.
\newblock \showarticletitle{On the use of automated text summarization
  techniques for summarizing source code}. In \bibinfo{booktitle}{\emph{2010
  17th Working Conference on Reverse Engineering}}. IEEE,
  \bibinfo{pages}{35--44}.
\newblock


\bibitem[He et~al\mbox{.}(2013)]%
        {he2013learning}
\bibfield{author}{\bibinfo{person}{Zhimin He}, \bibinfo{person}{Fayola Peters},
  \bibinfo{person}{Tim Menzies}, {and} \bibinfo{person}{Ye Yang}.}
  \bibinfo{year}{2013}\natexlab{}.
\newblock \showarticletitle{Learning from open-source projects: An empirical
  study on defect prediction}. In \bibinfo{booktitle}{\emph{2013 ACM/IEEE
  international symposium on empirical software engineering and measurement}}.
  IEEE, \bibinfo{pages}{45--54}.
\newblock


\bibitem[Hellendoorn and Devanbu(2017)]%
        {hellendoorn2017deep}
\bibfield{author}{\bibinfo{person}{Vincent~J Hellendoorn} {and}
  \bibinfo{person}{Premkumar Devanbu}.} \bibinfo{year}{2017}\natexlab{}.
\newblock \showarticletitle{Are deep neural networks the best choice for
  modeling source code?}. In \bibinfo{booktitle}{\emph{Proceedings of the 2017
  11th Joint meeting on foundations of software engineering}}.
  \bibinfo{pages}{763--773}.
\newblock


\bibitem[Hellendoorn et~al\mbox{.}(2019)]%
        {hellendoorn2019global}
\bibfield{author}{\bibinfo{person}{Vincent~J Hellendoorn},
  \bibinfo{person}{Charles Sutton}, \bibinfo{person}{Rishabh Singh},
  \bibinfo{person}{Petros Maniatis}, {and} \bibinfo{person}{David Bieber}.}
  \bibinfo{year}{2019}\natexlab{}.
\newblock \showarticletitle{Global relational models of source code}. In
  \bibinfo{booktitle}{\emph{International conference on learning
  representations}}.
\newblock


\bibitem[Hindle et~al\mbox{.}(2012)]%
        {hindle2012naturalness}
\bibfield{author}{\bibinfo{person}{Abram Hindle}, \bibinfo{person}{Earl~T
  Barr}, \bibinfo{person}{Zhendong Su}, \bibinfo{person}{Mark Gabel}, {and}
  \bibinfo{person}{Premkumar Devanbu}.} \bibinfo{year}{2012}\natexlab{}.
\newblock \bibinfo{title}{On the naturalness of software. In2012 34th
  International Conference on Software Engineering (ICSE)}.
\newblock
\newblock


\bibitem[Hosseini et~al\mbox{.}(2017)]%
        {hosseini2017systematic}
\bibfield{author}{\bibinfo{person}{Seyedrebvar Hosseini},
  \bibinfo{person}{Burak Turhan}, {and} \bibinfo{person}{Dimuthu Gunarathna}.}
  \bibinfo{year}{2017}\natexlab{}.
\newblock \showarticletitle{A systematic literature review and meta-analysis on
  cross project defect prediction}.
\newblock \bibinfo{journal}{\emph{IEEE Transactions on Software Engineering}}
  \bibinfo{volume}{45}, \bibinfo{number}{2} (\bibinfo{year}{2017}),
  \bibinfo{pages}{111--147}.
\newblock


\bibitem[Hou et~al\mbox{.}(2023)]%
        {hou2023large}
\bibfield{author}{\bibinfo{person}{Xinyi Hou}, \bibinfo{person}{Yanjie Zhao},
  \bibinfo{person}{Yue Liu}, \bibinfo{person}{Zhou Yang},
  \bibinfo{person}{Kailong Wang}, \bibinfo{person}{Li Li},
  \bibinfo{person}{Xiapu Luo}, \bibinfo{person}{David Lo},
  \bibinfo{person}{John Grundy}, {and} \bibinfo{person}{Haoyu Wang}.}
  \bibinfo{year}{2023}\natexlab{}.
\newblock \showarticletitle{Large language models for software engineering: A
  systematic literature review}.
\newblock \bibinfo{journal}{\emph{arXiv preprint arXiv:2308.10620}}
  (\bibinfo{year}{2023}).
\newblock


\bibitem[Hu et~al\mbox{.}(2018)]%
        {hu2018summarizing}
\bibfield{author}{\bibinfo{person}{Xing Hu}, \bibinfo{person}{Ge Li},
  \bibinfo{person}{Xin Xia}, \bibinfo{person}{David Lo}, \bibinfo{person}{Shuai
  Lu}, {and} \bibinfo{person}{Zhi Jin}.} \bibinfo{year}{2018}\natexlab{}.
\newblock \showarticletitle{Summarizing source code with transferred API
  knowledge}. In \bibinfo{booktitle}{\emph{Proceedings of the 27th
  International Joint Conference on Artificial Intelligence}}.
  \bibinfo{pages}{2269--2275}.
\newblock


\bibitem[Iyer et~al\mbox{.}(2016)]%
        {iyer2016summarizing}
\bibfield{author}{\bibinfo{person}{Srinivasan Iyer}, \bibinfo{person}{Ioannis
  Konstas}, \bibinfo{person}{Alvin Cheung}, {and} \bibinfo{person}{Luke
  Zettlemoyer}.} \bibinfo{year}{2016}\natexlab{}.
\newblock \showarticletitle{Summarizing source code using a neural attention
  model}. In \bibinfo{booktitle}{\emph{Proceedings of the 54th Annual Meeting
  of the Association for Computational Linguistics (Volume 1: Long Papers)}}.
  \bibinfo{pages}{2073--2083}.
\newblock


\bibitem[Jesse et~al\mbox{.}(2023)]%
        {jesse2023large}
\bibfield{author}{\bibinfo{person}{Kevin Jesse}, \bibinfo{person}{Toufique
  Ahmed}, \bibinfo{person}{Premkumar~T Devanbu}, {and} \bibinfo{person}{Emily
  Morgan}.} \bibinfo{year}{2023}\natexlab{}.
\newblock \showarticletitle{Large Language Models and Simple, Stupid Bugs}.
\newblock \bibinfo{journal}{\emph{arXiv preprint arXiv:2303.11455}}
  (\bibinfo{year}{2023}).
\newblock


\bibitem[Jiang et~al\mbox{.}(2023)]%
        {jiang2023impact}
\bibfield{author}{\bibinfo{person}{Nan Jiang}, \bibinfo{person}{Kevin Liu},
  \bibinfo{person}{Thibaud Lutellier}, {and} \bibinfo{person}{Lin Tan}.}
  \bibinfo{year}{2023}\natexlab{}.
\newblock \showarticletitle{Impact of Code Language Models on Automated Program
  Repair}.
\newblock \bibinfo{journal}{\emph{ICSE}} (\bibinfo{year}{2023}).
\newblock


\bibitem[Kang et~al\mbox{.}(2023)]%
        {kang2023large}
\bibfield{author}{\bibinfo{person}{Sungmin Kang}, \bibinfo{person}{Juyeon
  Yoon}, {and} \bibinfo{person}{Shin Yoo}.} \bibinfo{year}{2023}\natexlab{}.
\newblock \showarticletitle{Large Language Models are Few-shot Testers:
  Exploring LLM-based General Bug Reproduction}.
\newblock \bibinfo{journal}{\emph{ICSE}} (\bibinfo{year}{2023}).
\newblock


\bibitem[Karampatsis et~al\mbox{.}(2020)]%
        {karampatsis2020big}
\bibfield{author}{\bibinfo{person}{Rafael-Michael Karampatsis},
  \bibinfo{person}{Hlib Babii}, \bibinfo{person}{Romain Robbes},
  \bibinfo{person}{Charles Sutton}, {and} \bibinfo{person}{Andrea Janes}.}
  \bibinfo{year}{2020}\natexlab{}.
\newblock \showarticletitle{Big code!= big vocabulary: Open-vocabulary models
  for source code}. In \bibinfo{booktitle}{\emph{Proceedings of the ACM/IEEE
  42nd International Conference on Software Engineering}}.
  \bibinfo{pages}{1073--1085}.
\newblock


\bibitem[LeClair et~al\mbox{.}(2019)]%
        {leclair2019neural}
\bibfield{author}{\bibinfo{person}{Alexander LeClair}, \bibinfo{person}{Siyuan
  Jiang}, {and} \bibinfo{person}{Collin McMillan}.}
  \bibinfo{year}{2019}\natexlab{}.
\newblock \showarticletitle{A neural model for generating natural language
  summaries of program subroutines}. In \bibinfo{booktitle}{\emph{2019 IEEE/ACM
  41st International Conference on Software Engineering (ICSE)}}. IEEE,
  \bibinfo{pages}{795--806}.
\newblock


\bibitem[Levenshtein et~al\mbox{.}(1966)]%
        {levenshtein1966binary}
\bibfield{author}{\bibinfo{person}{Vladimir~I Levenshtein} {et~al\mbox{.}}}
  \bibinfo{year}{1966}\natexlab{}.
\newblock \showarticletitle{Binary codes capable of correcting deletions,
  insertions, and reversals}. In \bibinfo{booktitle}{\emph{Soviet physics
  doklady}}, Vol.~\bibinfo{volume}{10}. Soviet Union,
  \bibinfo{pages}{707--710}.
\newblock


\bibitem[Lin(2004)]%
        {lin2004rouge}
\bibfield{author}{\bibinfo{person}{Chin-Yew Lin}.}
  \bibinfo{year}{2004}\natexlab{}.
\newblock \showarticletitle{Rouge: A package for automatic evaluation of
  summaries}. In \bibinfo{booktitle}{\emph{Text summarization branches out}}.
  \bibinfo{pages}{74--81}.
\newblock


\bibitem[Lin and Och(2004)]%
        {lin2004orange}
\bibfield{author}{\bibinfo{person}{Chin-Yew Lin} {and}
  \bibinfo{person}{Franz~Josef Och}.} \bibinfo{year}{2004}\natexlab{}.
\newblock \showarticletitle{Orange: a method for evaluating automatic
  evaluation metrics for machine translation}. In
  \bibinfo{booktitle}{\emph{COLING 2004: Proceedings of the 20th International
  Conference on Computational Linguistics}}. \bibinfo{pages}{501--507}.
\newblock


\bibitem[Liu et~al\mbox{.}(2023)]%
        {liu2023your}
\bibfield{author}{\bibinfo{person}{Jiawei Liu}, \bibinfo{person}{Chunqiu~Steven
  Xia}, \bibinfo{person}{Yuyao Wang}, {and} \bibinfo{person}{Lingming Zhang}.}
  \bibinfo{year}{2023}\natexlab{}.
\newblock \showarticletitle{Is your code generated by chatgpt really correct?
  rigorous evaluation of large language models for code generation}.
\newblock \bibinfo{journal}{\emph{arXiv preprint arXiv:2305.01210}}
  (\bibinfo{year}{2023}).
\newblock


\bibitem[Lu et~al\mbox{.}(2021)]%
        {lu2021codexglue}
\bibfield{author}{\bibinfo{person}{Shuai Lu}, \bibinfo{person}{Daya Guo},
  \bibinfo{person}{Shuo Ren}, \bibinfo{person}{Junjie Huang},
  \bibinfo{person}{Alexey Svyatkovskiy}, \bibinfo{person}{Ambrosio Blanco},
  \bibinfo{person}{Colin Clement}, \bibinfo{person}{Dawn Drain},
  \bibinfo{person}{Daxin Jiang}, \bibinfo{person}{Duyu Tang}, {et~al\mbox{.}}}
  \bibinfo{year}{2021}\natexlab{}.
\newblock \showarticletitle{Codexglue: A machine learning benchmark dataset for
  code understanding and generation}.
\newblock \bibinfo{journal}{\emph{arXiv preprint arXiv:2102.04664}}
  (\bibinfo{year}{2021}).
\newblock


\bibitem[Meyerovich and Rabkin(2013)]%
        {meyerovich2013empirical}
\bibfield{author}{\bibinfo{person}{Leo~A Meyerovich} {and}
  \bibinfo{person}{Ariel~S Rabkin}.} \bibinfo{year}{2013}\natexlab{}.
\newblock \showarticletitle{Empirical analysis of programming language
  adoption}. In \bibinfo{booktitle}{\emph{Proceedings of the 2013 ACM SIGPLAN
  international conference on Object oriented programming systems languages \&
  applications}}. \bibinfo{pages}{1--18}.
\newblock


\bibitem[Murali et~al\mbox{.}(2023)]%
        {murali2023codecompose}
\bibfield{author}{\bibinfo{person}{Vijayaraghavan Murali},
  \bibinfo{person}{Chandra Maddila}, \bibinfo{person}{Imad Ahmad},
  \bibinfo{person}{Michael Bolin}, \bibinfo{person}{Daniel Cheng},
  \bibinfo{person}{Negar Ghorbani}, \bibinfo{person}{Renuka Fernandez}, {and}
  \bibinfo{person}{Nachiappan Nagappan}.} \bibinfo{year}{2023}\natexlab{}.
\newblock \showarticletitle{CodeCompose: A Large-Scale Industrial Deployment of
  AI-assisted Code Authoring}.
\newblock \bibinfo{journal}{\emph{arXiv preprint arXiv:2305.12050}}
  (\bibinfo{year}{2023}).
\newblock


\bibitem[Nashid et~al\mbox{.}(2023)]%
        {nashid2023retrieval}
\bibfield{author}{\bibinfo{person}{Noor Nashid}, \bibinfo{person}{Mifta
  Sintaha}, {and} \bibinfo{person}{Ali Mesbah}.}
  \bibinfo{year}{2023}\natexlab{}.
\newblock \showarticletitle{Retrieval-Based Prompt Selection for Code-Related
  Few-Shot Learning}. In \bibinfo{booktitle}{\emph{Proceedings, 45th ICSE}}.
\newblock


\bibitem[Nijkamp et~al\mbox{.}(2023)]%
        {nijkamp2023codegen2}
\bibfield{author}{\bibinfo{person}{Erik Nijkamp}, \bibinfo{person}{Hiroaki
  Hayashi}, \bibinfo{person}{Caiming Xiong}, \bibinfo{person}{Silvio Savarese},
  {and} \bibinfo{person}{Yingbo Zhou}.} \bibinfo{year}{2023}\natexlab{}.
\newblock \showarticletitle{Codegen2: Lessons for training llms on programming
  and natural languages}.
\newblock \bibinfo{journal}{\emph{arXiv preprint arXiv:2305.02309}}
  (\bibinfo{year}{2023}).
\newblock


\bibitem[Nijkamp et~al\mbox{.}(2022)]%
        {nijkamp2022codegen}
\bibfield{author}{\bibinfo{person}{Erik Nijkamp}, \bibinfo{person}{Bo Pang},
  \bibinfo{person}{Hiroaki Hayashi}, \bibinfo{person}{Lifu Tu},
  \bibinfo{person}{Huan Wang}, \bibinfo{person}{Yingbo Zhou},
  \bibinfo{person}{Silvio Savarese}, {and} \bibinfo{person}{Caiming Xiong}.}
  \bibinfo{year}{2022}\natexlab{}.
\newblock \showarticletitle{Codegen: An open large language model for code with
  multi-turn program synthesis}.
\newblock \bibinfo{journal}{\emph{arXiv preprint arXiv:2203.13474}}
  (\bibinfo{year}{2022}).
\newblock


\bibitem[Octoverse({[n.\,d.]})]%
        {githubProgrammingLanguages}
\bibfield{author}{\bibinfo{person}{Github Octoverse}.}
  \bibinfo{year}{[n.\,d.]}\natexlab{}.
\newblock \bibinfo{title}{{T}he top programming languages ---
  octoverse.github.com}.
\newblock
  \bibinfo{howpublished}{\url{https://octoverse.github.com/2022/top-programming-languages}}.
\newblock
\newblock
\shownote{[Accessed 22-02-2024]}.


\bibitem[Ouyang et~al\mbox{.}(2022)]%
        {ouyang2022training}
\bibfield{author}{\bibinfo{person}{Long Ouyang}, \bibinfo{person}{Jeffrey Wu},
  \bibinfo{person}{Xu Jiang}, \bibinfo{person}{Diogo Almeida},
  \bibinfo{person}{Carroll Wainwright}, \bibinfo{person}{Pamela Mishkin},
  \bibinfo{person}{Chong Zhang}, \bibinfo{person}{Sandhini Agarwal},
  \bibinfo{person}{Katarina Slama}, \bibinfo{person}{Alex Ray},
  {et~al\mbox{.}}} \bibinfo{year}{2022}\natexlab{}.
\newblock \showarticletitle{Training language models to follow instructions
  with human feedback}.
\newblock \bibinfo{journal}{\emph{Advances in Neural Information Processing
  Systems}}  \bibinfo{volume}{35} (\bibinfo{year}{2022}),
  \bibinfo{pages}{27730--27744}.
\newblock


\bibitem[Pal and Sillitti(2022)]%
        {pal2022cross}
\bibfield{author}{\bibinfo{person}{Sourabh Pal} {and} \bibinfo{person}{Alberto
  Sillitti}.} \bibinfo{year}{2022}\natexlab{}.
\newblock \showarticletitle{Cross-Project Defect Prediction: A Literature
  Review}.
\newblock \bibinfo{journal}{\emph{IEEE Access}} (\bibinfo{year}{2022}).
\newblock


\bibitem[Papineni et~al\mbox{.}(2002)]%
        {papineni2002bleu}
\bibfield{author}{\bibinfo{person}{Kishore Papineni}, \bibinfo{person}{Salim
  Roukos}, \bibinfo{person}{Todd Ward}, {and} \bibinfo{person}{Wei-Jing Zhu}.}
  \bibinfo{year}{2002}\natexlab{}.
\newblock \showarticletitle{Bleu: a method for automatic evaluation of machine
  translation}. In \bibinfo{booktitle}{\emph{Proceedings of the 40th annual
  meeting of the Association for Computational Linguistics}}.
  \bibinfo{pages}{311--318}.
\newblock


\bibitem[Paulson et~al\mbox{.}(2004)]%
        {paulson2004empirical}
\bibfield{author}{\bibinfo{person}{James~W Paulson}, \bibinfo{person}{Giancarlo
  Succi}, {and} \bibinfo{person}{Armin Eberlein}.}
  \bibinfo{year}{2004}\natexlab{}.
\newblock \showarticletitle{An empirical study of open-source and closed-source
  software products}.
\newblock \bibinfo{journal}{\emph{IEEE transactions on software engineering}}
  \bibinfo{volume}{30}, \bibinfo{number}{4} (\bibinfo{year}{2004}),
  \bibinfo{pages}{246--256}.
\newblock


\bibitem[Pradel et~al\mbox{.}(2020)]%
        {pradel2020typewriter}
\bibfield{author}{\bibinfo{person}{Michael Pradel}, \bibinfo{person}{Georgios
  Gousios}, \bibinfo{person}{Jason Liu}, {and} \bibinfo{person}{Satish
  Chandra}.} \bibinfo{year}{2020}\natexlab{}.
\newblock \showarticletitle{Typewriter: Neural type prediction with
  search-based validation}. In \bibinfo{booktitle}{\emph{Proceedings of the
  28th ACM Joint Meeting on European Software Engineering Conference and
  Symposium on the Foundations of Software Engineering}}.
  \bibinfo{pages}{209--220}.
\newblock


\bibitem[Rabin et~al\mbox{.}(2023)]%
        {rabin2023memorization}
\bibfield{author}{\bibinfo{person}{Md~Rafiqul~Islam Rabin},
  \bibinfo{person}{Aftab Hussain}, \bibinfo{person}{Mohammad~Amin Alipour},
  {and} \bibinfo{person}{Vincent~J Hellendoorn}.}
  \bibinfo{year}{2023}\natexlab{}.
\newblock \showarticletitle{Memorization and generalization in neural code
  intelligence models}.
\newblock \bibinfo{journal}{\emph{Information and Software Technology}}
  \bibinfo{volume}{153} (\bibinfo{year}{2023}), \bibinfo{pages}{107066}.
\newblock


\bibitem[Radford et~al\mbox{.}(2018)]%
        {radford2018improving}
\bibfield{author}{\bibinfo{person}{Alec Radford}, \bibinfo{person}{Karthik
  Narasimhan}, \bibinfo{person}{Tim Salimans}, \bibinfo{person}{Ilya
  Sutskever}, {et~al\mbox{.}}} \bibinfo{year}{2018}\natexlab{}.
\newblock \showarticletitle{Improving language understanding by generative
  pre-training}.
\newblock  (\bibinfo{year}{2018}).
\newblock


\bibitem[Radford et~al\mbox{.}(2019)]%
        {radford2019language}
\bibfield{author}{\bibinfo{person}{Alec Radford}, \bibinfo{person}{Jeffrey Wu},
  \bibinfo{person}{Rewon Child}, \bibinfo{person}{David Luan},
  \bibinfo{person}{Dario Amodei}, \bibinfo{person}{Ilya Sutskever},
  {et~al\mbox{.}}} \bibinfo{year}{2019}\natexlab{}.
\newblock \showarticletitle{Language models are unsupervised multitask
  learners}.
\newblock \bibinfo{journal}{\emph{OpenAI blog}} \bibinfo{volume}{1},
  \bibinfo{number}{8} (\bibinfo{year}{2019}), \bibinfo{pages}{9}.
\newblock


\bibitem[Ramos et~al\mbox{.}(2003)]%
        {ramos2003using}
\bibfield{author}{\bibinfo{person}{Juan Ramos} {et~al\mbox{.}}}
  \bibinfo{year}{2003}\natexlab{}.
\newblock \showarticletitle{Using tf-idf to determine word relevance in
  document queries}. In \bibinfo{booktitle}{\emph{Proceedings of the first
  instructional conference on machine learning}}, Vol.~\bibinfo{volume}{242}.
  Citeseer, \bibinfo{pages}{29--48}.
\newblock


\bibitem[Ren et~al\mbox{.}(2020)]%
        {ren2020codebleu}
\bibfield{author}{\bibinfo{person}{Shuo Ren}, \bibinfo{person}{Daya Guo},
  \bibinfo{person}{Shuai Lu}, \bibinfo{person}{Long Zhou},
  \bibinfo{person}{Shujie Liu}, \bibinfo{person}{Duyu Tang},
  \bibinfo{person}{Neel Sundaresan}, \bibinfo{person}{Ming Zhou},
  \bibinfo{person}{Ambrosio Blanco}, {and} \bibinfo{person}{Shuai Ma}.}
  \bibinfo{year}{2020}\natexlab{}.
\newblock \showarticletitle{Codebleu: a method for automatic evaluation of code
  synthesis}.
\newblock \bibinfo{journal}{\emph{arXiv preprint arXiv:2009.10297}}
  (\bibinfo{year}{2020}).
\newblock


\bibitem[Robertson et~al\mbox{.}(2009)]%
        {robertson2009probabilistic}
\bibfield{author}{\bibinfo{person}{Stephen Robertson}, \bibinfo{person}{Hugo
  Zaragoza}, {et~al\mbox{.}}} \bibinfo{year}{2009}\natexlab{}.
\newblock \showarticletitle{The probabilistic relevance framework: BM25 and
  beyond}.
\newblock \bibinfo{journal}{\emph{Foundations and Trends{\textregistered} in
  Information Retrieval}} \bibinfo{volume}{3}, \bibinfo{number}{4}
  (\bibinfo{year}{2009}), \bibinfo{pages}{333--389}.
\newblock


\bibitem[Rodeghero et~al\mbox{.}(2014)]%
        {rodeghero2014improving}
\bibfield{author}{\bibinfo{person}{Paige Rodeghero}, \bibinfo{person}{Collin
  McMillan}, \bibinfo{person}{Paul~W McBurney}, \bibinfo{person}{Nigel Bosch},
  {and} \bibinfo{person}{Sidney D'Mello}.} \bibinfo{year}{2014}\natexlab{}.
\newblock \showarticletitle{Improving automated source code summarization via
  an eye-tracking study of programmers}. In
  \bibinfo{booktitle}{\emph{Proceedings of the 36th international conference on
  Software engineering}}. \bibinfo{pages}{390--401}.
\newblock


\bibitem[Roy et~al\mbox{.}(2021)]%
        {roy2021reassessing}
\bibfield{author}{\bibinfo{person}{Devjeet Roy}, \bibinfo{person}{Sarah
  Fakhoury}, {and} \bibinfo{person}{Venera Arnaoudova}.}
  \bibinfo{year}{2021}\natexlab{}.
\newblock \showarticletitle{Reassessing automatic evaluation metrics for code
  summarization tasks}. In \bibinfo{booktitle}{\emph{Proceedings of the 29th
  ACM Joint Meeting on European Software Engineering Conference and Symposium
  on the Foundations of Software Engineering}}. \bibinfo{pages}{1105--1116}.
\newblock


\bibitem[Roziere et~al\mbox{.}(2023)]%
        {roziere2023code}
\bibfield{author}{\bibinfo{person}{Baptiste Roziere}, \bibinfo{person}{Jonas
  Gehring}, \bibinfo{person}{Fabian Gloeckle}, \bibinfo{person}{Sten Sootla},
  \bibinfo{person}{Itai Gat}, \bibinfo{person}{Xiaoqing~Ellen Tan},
  \bibinfo{person}{Yossi Adi}, \bibinfo{person}{Jingyu Liu},
  \bibinfo{person}{Tal Remez}, \bibinfo{person}{J{\'e}r{\'e}my Rapin},
  {et~al\mbox{.}}} \bibinfo{year}{2023}\natexlab{}.
\newblock \showarticletitle{Code llama: Open foundation models for code}.
\newblock \bibinfo{journal}{\emph{arXiv preprint arXiv:2308.12950}}
  (\bibinfo{year}{2023}).
\newblock


\bibitem[Sanders({[n.\,d.]})]%
        {thinkfulCodingLanguages}
\bibfield{author}{\bibinfo{person}{Abby Sanders}.}
  \bibinfo{year}{[n.\,d.]}\natexlab{}.
\newblock \bibinfo{title}{{T}he {T}op {C}oding {L}anguages by {I}ndustry ---
  thinkful.com}.
\newblock
  \bibinfo{howpublished}{\url{https://www.thinkful.com/blog/top-coding-languages-by-industry/}}.
\newblock
\newblock
\shownote{[Accessed 22-02-2024]}.


\bibitem[Shi et~al\mbox{.}(2023)]%
        {shi2022evaluation}
\bibfield{author}{\bibinfo{person}{Ensheng Shi}, \bibinfo{person}{Yanlin Wang},
  \bibinfo{person}{Lun Du}, \bibinfo{person}{Junjie Chen}, \bibinfo{person}{Shi
  Han}, \bibinfo{person}{Hongyu Zhang}, \bibinfo{person}{Dongmei Zhang}, {and}
  \bibinfo{person}{Hongbin Sun}.} \bibinfo{year}{2023}\natexlab{}.
\newblock \showarticletitle{On the evaluation of neural code summarization}. In
  \bibinfo{booktitle}{\emph{Proceedings of the 44th International Conference on
  Software Engineering}}. \bibinfo{pages}{1597--1608}.
\newblock


\bibitem[Sridhara et~al\mbox{.}(2010)]%
        {sridhara2010towards}
\bibfield{author}{\bibinfo{person}{Giriprasad Sridhara}, \bibinfo{person}{Emily
  Hill}, \bibinfo{person}{Divya Muppaneni}, \bibinfo{person}{Lori Pollock},
  {and} \bibinfo{person}{K Vijay-Shanker}.} \bibinfo{year}{2010}\natexlab{}.
\newblock \showarticletitle{Towards automatically generating summary comments
  for java methods}. In \bibinfo{booktitle}{\emph{Proceedings of the IEEE/ACM
  international conference on Automated software engineering}}.
  \bibinfo{pages}{43--52}.
\newblock


\bibitem[Tabachnyk and Nikolov({[n.\,d.]})]%
        {tabachnyk2022ml}
\bibfield{author}{\bibinfo{person}{Maxim Tabachnyk} {and}
  \bibinfo{person}{Stoyan Nikolov}.} \bibinfo{year}{[n.\,d.]}\natexlab{}.
\newblock \showarticletitle{Ml-enhanced code completion improves developer
  productivity, 2022}.
\newblock \bibinfo{journal}{\emph{URL
  https://ai.googleblog.com/2022/07/ml-enhanced-code-completion-improves.html}}
  (\bibinfo{year}{[n.\,d.]}).
\newblock


\bibitem[Tu et~al\mbox{.}(2014)]%
        {tu2014localness}
\bibfield{author}{\bibinfo{person}{Zhaopeng Tu}, \bibinfo{person}{Zhendong Su},
  {and} \bibinfo{person}{Premkumar Devanbu}.} \bibinfo{year}{2014}\natexlab{}.
\newblock \showarticletitle{On the localness of software}. In
  \bibinfo{booktitle}{\emph{Proceedings of the 22nd ACM SIGSOFT International
  Symposium on Foundations of Software Engineering}}.
  \bibinfo{pages}{269--280}.
\newblock


\bibitem[Van~der Maaten and Hinton(2008)]%
        {van2008visualizing}
\bibfield{author}{\bibinfo{person}{Laurens Van~der Maaten} {and}
  \bibinfo{person}{Geoffrey Hinton}.} \bibinfo{year}{2008}\natexlab{}.
\newblock \showarticletitle{Visualizing data using t-SNE.}
\newblock \bibinfo{journal}{\emph{Journal of machine learning research}}
  \bibinfo{volume}{9}, \bibinfo{number}{11} (\bibinfo{year}{2008}).
\newblock


\bibitem[Vaswani et~al\mbox{.}(2017)]%
        {vaswani2017attention}
\bibfield{author}{\bibinfo{person}{Ashish Vaswani}, \bibinfo{person}{Noam
  Shazeer}, \bibinfo{person}{Niki Parmar}, \bibinfo{person}{Jakob Uszkoreit},
  \bibinfo{person}{Llion Jones}, \bibinfo{person}{Aidan~N Gomez},
  \bibinfo{person}{{\L}ukasz Kaiser}, {and} \bibinfo{person}{Illia
  Polosukhin}.} \bibinfo{year}{2017}\natexlab{}.
\newblock \showarticletitle{Attention is all you need}. In
  \bibinfo{booktitle}{\emph{Advances in neural information processing
  systems}}. \bibinfo{pages}{5998--6008}.
\newblock


\bibitem[Wang et~al\mbox{.}(2022)]%
        {wang2022self}
\bibfield{author}{\bibinfo{person}{Xuezhi Wang}, \bibinfo{person}{Jason Wei},
  \bibinfo{person}{Dale Schuurmans}, \bibinfo{person}{Quoc Le},
  \bibinfo{person}{Ed Chi}, \bibinfo{person}{Sharan Narang},
  \bibinfo{person}{Aakanksha Chowdhery}, {and} \bibinfo{person}{Denny Zhou}.}
  \bibinfo{year}{2022}\natexlab{}.
\newblock \showarticletitle{Self-consistency improves chain of thought
  reasoning in language models}.
\newblock \bibinfo{journal}{\emph{arXiv preprint arXiv:2203.11171}}
  (\bibinfo{year}{2022}).
\newblock


\bibitem[Wang et~al\mbox{.}(2020)]%
        {wang2020generalizing}
\bibfield{author}{\bibinfo{person}{Yaqing Wang}, \bibinfo{person}{Quanming
  Yao}, \bibinfo{person}{James~T Kwok}, {and} \bibinfo{person}{Lionel~M Ni}.}
  \bibinfo{year}{2020}\natexlab{}.
\newblock \showarticletitle{Generalizing from a few examples: A survey on
  few-shot learning}.
\newblock \bibinfo{journal}{\emph{ACM computing surveys (csur)}}
  \bibinfo{volume}{53}, \bibinfo{number}{3} (\bibinfo{year}{2020}),
  \bibinfo{pages}{1--34}.
\newblock


\bibitem[Wei et~al\mbox{.}(2022)]%
        {wei2022chain}
\bibfield{author}{\bibinfo{person}{Jason Wei}, \bibinfo{person}{Xuezhi Wang},
  \bibinfo{person}{Dale Schuurmans}, \bibinfo{person}{Maarten Bosma},
  \bibinfo{person}{Fei Xia}, \bibinfo{person}{Ed Chi}, \bibinfo{person}{Quoc~V
  Le}, \bibinfo{person}{Denny Zhou}, {et~al\mbox{.}}}
  \bibinfo{year}{2022}\natexlab{}.
\newblock \showarticletitle{Chain-of-thought prompting elicits reasoning in
  large language models}.
\newblock \bibinfo{journal}{\emph{Advances in Neural Information Processing
  Systems}}  \bibinfo{volume}{35} (\bibinfo{year}{2022}),
  \bibinfo{pages}{24824--24837}.
\newblock


\bibitem[Wei et~al\mbox{.}(2023)]%
        {wei2023copiloting}
\bibfield{author}{\bibinfo{person}{Yuxiang Wei},
  \bibinfo{person}{Chunqiu~Steven Xia}, {and} \bibinfo{person}{Lingming
  Zhang}.} \bibinfo{year}{2023}\natexlab{}.
\newblock \showarticletitle{Copiloting the copilots: Fusing large language
  models with completion engines for automated program repair}. In
  \bibinfo{booktitle}{\emph{Proceedings of the 31st ACM Joint European Software
  Engineering Conference and Symposium on the Foundations of Software
  Engineering}}. \bibinfo{pages}{172--184}.
\newblock


\bibitem[Wikipedia({[n.\,d.]})]%
        {wikipediaListSharp}
\bibfield{author}{\bibinfo{person}{Wikipedia}.}
  \bibinfo{year}{[n.\,d.]}\natexlab{}.
\newblock \bibinfo{title}{{L}ist of {C} {S}harp software - {W}ikipedia ---
  en.wikipedia.org}.
\newblock
  \bibinfo{howpublished}{\url{https://en.wikipedia.org/wiki/List_of_C_Sharp_software}}.
\newblock
\newblock
\shownote{[Accessed 22-02-2024]}.


\bibitem[Xu et~al\mbox{.}(2022)]%
        {xu2022systematic}
\bibfield{author}{\bibinfo{person}{Frank~F Xu}, \bibinfo{person}{Uri Alon},
  \bibinfo{person}{Graham Neubig}, {and} \bibinfo{person}{Vincent~Josua
  Hellendoorn}.} \bibinfo{year}{2022}\natexlab{}.
\newblock \showarticletitle{A systematic evaluation of large language models of
  code}. In \bibinfo{booktitle}{\emph{Proceedings of the 6th ACM SIGPLAN
  International Symposium on Machine Programming}}. \bibinfo{pages}{1--10}.
\newblock


\bibitem[Yu et~al\mbox{.}(2023)]%
        {yu2023codereval}
\bibfield{author}{\bibinfo{person}{Hao Yu}, \bibinfo{person}{Bo Shen},
  \bibinfo{person}{Dezhi Ran}, \bibinfo{person}{Jiaxin Zhang},
  \bibinfo{person}{Qi Zhang}, \bibinfo{person}{Yuchi Ma},
  \bibinfo{person}{Guangtai Liang}, \bibinfo{person}{Ying Li},
  \bibinfo{person}{Tao Xie}, {and} \bibinfo{person}{Qianxiang Wang}.}
  \bibinfo{year}{2023}\natexlab{}.
\newblock \showarticletitle{CoderEval: A Benchmark of Pragmatic Code Generation
  with Generative Pre-trained Models}.
\newblock \bibinfo{journal}{\emph{arXiv preprint arXiv:2302.00288}}
  (\bibinfo{year}{2023}).
\newblock


\bibitem[Ziegler et~al\mbox{.}(2022)]%
        {ziegler2022productivity}
\bibfield{author}{\bibinfo{person}{Albert Ziegler}, \bibinfo{person}{Eirini
  Kalliamvakou}, \bibinfo{person}{X~Alice Li}, \bibinfo{person}{Andrew Rice},
  \bibinfo{person}{Devon Rifkin}, \bibinfo{person}{Shawn Simister},
  \bibinfo{person}{Ganesh Sittampalam}, {and} \bibinfo{person}{Edward
  Aftandilian}.} \bibinfo{year}{2022}\natexlab{}.
\newblock \showarticletitle{Productivity assessment of neural code completion}.
  In \bibinfo{booktitle}{\emph{Proceedings of the 6th ACM SIGPLAN International
  Symposium on Machine Programming}}. \bibinfo{pages}{21--29}.
\newblock


\bibitem[Zimmermann et~al\mbox{.}(2009)]%
        {zimmermann2009cross}
\bibfield{author}{\bibinfo{person}{Thomas Zimmermann},
  \bibinfo{person}{Nachiappan Nagappan}, \bibinfo{person}{Harald Gall},
  \bibinfo{person}{Emanuel Giger}, {and} \bibinfo{person}{Brendan Murphy}.}
  \bibinfo{year}{2009}\natexlab{}.
\newblock \showarticletitle{Cross-project defect prediction: a large scale
  experiment on data vs. domain vs. process}. In
  \bibinfo{booktitle}{\emph{Proceedings of the 7th joint meeting of the
  European software engineering conference and the ACM SIGSOFT symposium on The
  foundations of software engineering}}. \bibinfo{pages}{91--100}.
\newblock


\end{thebibliography}

\end{document}